\documentclass[12pt,preprint]{aastex}
\usepackage{appendix}
\begin{document}

\title{The Nature of Star Formation at 24\micron~in the Group Environment 
at 0.3 $\la$ z $\la$ 0.55}

\author{K. D. Tyler$^{1}$, G. H. Rieke$^{1}$, D. J. Wilman$^{2}$, S. L. McGee$^{3,4}$,
R. G. Bower$^{4}$, L. Bai,$^{5}$, J. S. Mulchaey$^{5}$, L. C. Parker$^{6}$, Y. Shi$^{1}$
D. Pierini$^{2}$}

\altaffiltext{1}{Steward Observatory, University of Arizona, 933 N. Cherry Ave.,
Tucson, AZ 85721, USA}
\altaffiltext{2}{Max-Planck-Institut f\"{u}r extraterrestrische Physik, Giessenbachstra{\ss}e, 
D-85748 Garching, Germany}
\altaffiltext{3}{Department of Physics and Astronomy, University of Waterloo, Waterloo, 
Ontario, N2L 3G1, Canada}
\altaffiltext{4}{Institute for Computational Cosmology, Department of Physics, Durham 
University, South Road, Durham DH1 3LE, UK}
\altaffiltext{5}{Observatories of the Carnegie Institution, 813 Santa Barbara Street, Pasadena,
CA, USA}
\altaffiltext{6}{Department of Physics and Astronomy, McMaster University, 1280 Main Street West, 
Hamilton, Ontario, L8S 4M1, Canada}

\begin{abstract}

Galaxy star formation rates (SFRs) are sensitive to the local environment; for example,
the high-density regions at the cores of dense clusters are known to 
suppress star formation.  It has been suggested that galaxy transformation occurs
largely in groups, which are the intermediate step in density between field and 
cluster environments.  
In this paper, we use deep MIPS 24\micron~observations 
of intermediate-redshift (0.3 $\la$ z $\la$ 0.55) group and field galaxies from the
Group Environment and Evolution Collaboration (GEEC) 
subset of the Second Canadian Network for Observational Cosmology (CNOC2) 
survey to probe the moderate-density environment of groups, 
wherein the majority of galaxies are found.  The completeness limit 
of our study is log(L$_{TIR}$(L$_{\odot}$)) $\ga$ 10.5, 
corresponding to SFR $\ga$ 2.7 M$_{\odot}$ yr$^{-1}$.  We find that the 
group and field galaxies have different distributions of morphologies and mass.
However, individual group galaxies have star-forming properties comparable 
to those of field galaxies
of similar mass and morphology; that is, the group environment does not appear to modify
the properties of these galaxies directly.  There is a relatively large number of massive 
early-type group spirals, along with E/S0 galaxies, 
that are forming stars above our detection limit.  
These galaxies account for the nearly comparable level of star-forming 
activity in groups as compared with the field, despite the differences in mass 
and morphology distributions between the two environments.
The distribution of specific SFRs
(SFR/M$_{*}$) is shifted to lower values in the groups, reflecting the fact that groups
contain a higher proportion of massive and less active galaxies. 
Considering the distributions of morphology, mass, and SFR, the group
members appear to lie between field and cluster galaxies in overall properties.

\end{abstract}



\section{INTRODUCTION}

In a $\Lambda$CDM universe, galaxies, on average, move from areas of lower density to areas of higher 
density, merging and combining to form larger and larger systems like the 
massive galaxy clusters we see in the local universe. Galaxy evolution, therefore, cannot be 
understood without considering the influence of environment on galaxy properties.
The importance of environment on galaxy evolution is demonstrated by the 
differences between galaxies in clusters and those in the field. For example,  
the fraction of blue galaxies in clusters has been 
decreasing since z $\sim$ 1, and local clusters are dominated by red galaxies 
(Butcher \& Oemler 1978; Kennicutt 1983; Hashimoto et al. 1998; Andreon et al. 2004; 
Poggianti et al. 2006; Cooper et al. 2007; Loh et al. 2008; Cucciati et al. 2010). 
Dressler (1980) found a dramatic increase in the 
proportion of early-type galaxies with local density inside rich clusters, 
i.e., the morphology-density relation.
This relation has also been shown to extend down to the group 
environment (Postman \& Geller 1984). Most of these cluster early-type galaxies have not had 
appreciable star formation in gigayears, though as we move outward from the centers of the clusters, 
we see more late-type galaxies overall and more early-type galaxies that have had more recent star 
formation (Balogh et al. 1997, 1999; Bai et al. 2009).

Environment is expected to influence the rates of both gas exhaustion and interactions.  The 
drop in star-forming activity in dense environments might be due to interactions with the 
inter-galactic gas, such as through ram-pressure stripping of cold gas in galaxies (Gunn \& Gott 
1972; Larson et al. 1980; Kinney et al. 2004) or stripping of the hot gas 
through strangulation (Balogh et al. 2000; Kawata \& Mulchaey 2008; McCarthy et al. 
2008). Alternatively, the lower fraction of star forming galaxies might 
arise through galaxy-galaxy interactions, either major mergers 
or harassment (frequent high-speed encounters of galaxies that do not lead to mergers), both of 
which would accelerate the star formation and lead to early exhaustion of the interstellar 
material. Tides raised by the overall gravitational potential of dense clusters may also play a role 
(Henriksen \& Byrd 1996). It has been proposed that the morphological transformation into 
early-type galaxies may occur first, suppressing star formation by stabilizing the gas against
fragmentation (Martig et al. 2009), although other recent studies 
question this claim \citep{kovac10}.

Explaining the behavior of galaxies in different environments in terms of consistent 
theories for the growth of galaxies in the early 
universe has proven challenging (Bower et al. 2006; Kaviraj et al. 2009). Therefore, an intense 
observational approach is needed to help develop an understanding of the relation between 
environment and galaxy evolution. It should be possible to disentangle environmental influences 
by comparing galaxy behavior in different environments, though just as the 
mechanisms for galaxy transformation are not fully understood, 
the environments where these processes operate also need to be explored and defined. 
Additionally, most of the previous work regarding star formation with respect to 
environment focused on rich clusters (e.g., Dressler et al. 2009; Bai et al. 2009; 
Haines et al. 2009, and references therein); 
relatively few studies have focused on groups (e.g. Zabludoff \& Mulchaey 1998, 2000; 
Balogh et al. 2004; Johnson et al. 2007; Marcillac et al. 2008; Bai et al. 2010).

Simard et al. (2009) argue that cluster-centric processes are not 
the dominant factor in galaxy morphological transformation.  The majority of 
galaxies live in the less-dense group environment (Geller \& Huchra 1983; Eke et al 2004),  
and because clusters probably form from coalescing groups and field galaxies, much of the evolution 
apparent in cluster galaxies may have occurred in groups prior to their assimilation 
into clusters. Zabludoff \& Mulchaey (1998) 
found that the proportion of early-type galaxies in groups ranged from that typical of the field 
($\sim$25\%) to that found in dense clusters ($\sim$55\%), suggesting that much of the 
morphological transformation of galaxies from field to cluster properties occurs in groups. 
\citet{just10} show that an increase in the proportion of S0 
galaxies with decreasing redshift occurs in moderate-mass groups/poor 
clusters ($\sigma < 750$ km s$^{-1}$), and \citet{wilman09} find that for groups at
intermediate redshifts, the fraction of S0s is as high as in clusters, even at fixed luminosities.  
Additionally, because most galaxies reside in groups---and, therefore, galaxies 
spend most of their time in groups---uncovering the effects of these lower-density environments 
can help us understand the evolutionary path of the global galaxy population over
cosmic time \citep{mcgee09}.

If morphological transformations frequently occur at intermediate densities, 
is this also the location where
star formation is cutoff? Previous studies of star formation in groups have mostly focused on 
optical indicators such as H$\alpha$ or [OII]$\lambda$3727 emission lines or ultraviolet 
continuum. These measures can extend to low levels of star formation, and they indicate a 
significantly lower level of activity in groups and clusters than in the field
(Wilman et al. 2005a; Gerke et al. 2007; Balogh et al. 2009; Iovino et al. 2010; Peng et al. 2010). 
Such first-order comparisons are usually made under the assumption that the extinction
is similar in groups and clusters (and that star formation is not deeply obscured).
However, corrections are required to convert these optical indicators to accurate 
star formation rates (SFRs).  At z $\ga$ 0.3,
H$\alpha$ moves out of the range of optical spectroscopy and SFR estimates rely on [OII],
where the extinction is large and uncertain.
The [OII]$\lambda3727$ line has additional problems of being 
sensitive to dust reddening and metallicity, thus being relatively weak in high-mass 
systems, although empirical corrections have been suggested to correct for this 
and other effects (Moustakas et al. 2006; Gilbank et al. 2010). 

The infrared (IR) is advantageous for probing high levels of star formation. 
While SFRs determined in the IR are only 
sensitive to dust-obscured star formation, the correction to the total star formation in objects 
at moderate to high luminosities (L$_{TIR} \ga 1 \times 10^{10}$ L$_{\odot}$) is small 
\citep{rieke09}. Nonetheless, previous IR studies have not reached consistent conclusions about 
star formation in groups.  \citet{marcillac08} found no significant 
dependence on the incidence of luminous infrared galaxies (LIRGs) as a function of field or group 
environment at z $\sim$ 0.8.  At lower redshifts, 
\citet{wilman08} find a dearth of star formation in group galaxies 
at z $\sim$ 0.4, while \citet{tran09} find a similar incidence of SFRs 
in massive groups (for galaxies forming stars $> 3 M_\odot$ yr$^{-1}$) as in the field
at z $\sim$ 0.37.  For local groups, Bai et al. (2010) report rates of star formation 
somewhat lower than in the field (by $\sim$30\%). 

This paper is a step toward understanding the influence of the group environment on star 
formation and resolving some of the apparent discrepancies in previous studies of the 
same topic.  Some of these discrepancies, especially in cluster studies, may arise from ``field''
samples contaminated by groups.  To compare groups with actual field galaxies, we need a 
clean field sample with as many galaxies as possible.
Here, we present 24\micron~measurements of 232 group galaxies and 236 field galaxies from 
0.3 $\la$ z $\la$ 0.55 in the Second Canadian Network for Observational Cosmology (CNOC2) survey.

The paper is organized as follows. In Section 2, we discuss the sample, data reductions, and 
errors.  Section 3 covers the construction of fractional IR luminosity functions (LFs) and compares the IR 
luminosities, morphologies, and masses of group and field galaxies to estimate what effect (if any) 
environment has on star formation at these redshifts.  We discuss the implications of our 
results in Section 4. For all cosmological corrections, we assume the parameters $H_0 = 
$70~km~s$^{-1}$Mpc$^{-1}$,~$\Omega_M = 0.3,~\Omega_\Lambda = 0.7$.

\section{SAMPLE AND DATA REDUCTION}  

\subsection{Sample Selection and Photometry}  

The CNOC2 survey is a photometric and spectroscopic survey of faint galaxies covering 
more than 1.5 deg$^2$ over four widely-spaced patches of sky (Yee et al. 2000; Carlberg 
et al. 2001a).  The original survey included five-color photometry in
$I_C$, $R_C$, V, B, and U to $R_C \sim 23.0$ (Vega) mag and spectroscopic redshifts (to $R_C \sim
21.5$ mag) for an unbiased sample of $\sim$6000 galaxies with the purpose of studying
galaxy clustering, dynamics and evolution at intermediate redshifts ($0.1 \la z \la 0.6$).
The survey is spectroscopically incomplete, but the selection function is very well defined
for this redshift range (Lin et al. 1999; Yee et al. 2000).

The groups themselves were originally selected by using a friends-of-friends algorithm to find
overdensities of galaxies in three-dimensional space (Carlberg et al. 2001a, 2001b).  
Follow-up LDSS2 spectroscopy (to $R_C \sim$ 22.0) targeted 26 of these Carlberg groups 
at $0.3 \la z \la 0.55$, creating a sample
of group and field galaxies to a greater depth than the original sample, 
with increased and unbiased spectroscopic completeness \citep{wilman05b}.  
Additional group members were carefully selected to ensure that the resulting sample
would be unbiased with regard to color using the method outlined in \citet{wilman05}.
This subset of the original CNOC2
survey---26 groups at 0.3 $\la$ z $\la$ 0.55 to $R_C \sim$ 22.0---was followed up 
by the Group Environment and Evolution Collaboration (GEEC), and hereafter, we refer
to these as the GEEC groups. 
Visual \citep{wilman09} and automated \citep{mcgee08} 
morphological classifications were made using deep, high-resolution
{\it Hubble Space Telescope} (HST) Advanced Camera for Surveys (ACS) 
images of the same 26 groups (Wilman et al. 2009).  Additional multiwavelength
coverage includes {\it Galaxy Evolution Explorer} (GALEX) UV \citep{mcgee}, and IRAC 
(Balogh et al. 2007, 2009; Wilman et al. 2008).  X-ray properties of a subset of
the groups have been measured with {\it XMM-Newton} and {\it Chandra} observations \citep{finoguenov09}.

We present observations of these GEEC groups (as defined by \citet{wilman05}) using the MIPS 
24\micron~band on the {\it Spitzer Space Telescope} \citep{rieke04}.  We used MIPS 
in single-source photometry mode (field of 5$^\prime$ $\times$ 5$^\prime$) due to the compact nature of the
groups, and long exposure times allowed us to detect relatively low SFRs 
($<$ 10 M$_{\odot}$ yr$^{-1}$).  All but five of the groups have either complete coverage
of their members with MIPS or are only missing one or two member galaxies.  All
galaxies assigned to groups using the algorithm described by \citet{wilman05} and within
the MIPS field of view are considered group members in this paper.  The MIPS 
24\micron~field of view corresponds to $\sim$1.3 Mpc ($\sim$2 Mpc) on a side at z $=$ 0.3 (0.55). 

The data from these observations were reduced using version 3.10 of the MIPS Data Analysis
Tool (DAT; Gordon et al. 2007).  Fields with overlapping regions were mosaicked for better
coverage near the image edges.  Sources $3\sigma$ above the standard deviation of the 
background were identified using DAOFIND in the 
IRAF\footnote[2]{IRAF is distributed and supported by the National Optical Astronomy 
Observatories (NOAO).} environment.  Flux densities were calculated using a point-spread
function (PSF) made from
bright sources in one of the larger mosaicked fields and the IRAF PSF-fitting routine 
ALLSTAR, correcting for flux lost in the wings of the PSF as described by the 
Spitzer Science Center\footnote[3]{The Spitzer Science Center (http://ssc.spitzer.caltech.edu)
is supported by NASA, the Jet Propulsion Laboratory, the California Institute
of Technology, and the Infrared Processing and Analysis Center.}.  

Our initial field sample consisted of all galaxies not identified
as residing in a group.  Given the completeness of the group sample, it has been estimated
that $\sim$21\% of this field sample is contaminated by unidentified groups
(Carlberg et al. 2001a, McGee et al. 2008).
To reduce the amount of group contamination in the field, we plotted a 
running-average histogram of redshifts for the LDSS2 pencil-beam fields with a bin size of 
z $=$ 0.001.  Any galaxies that fall in a bin with five or more galaxies at any point in the 
running average were removed from the field sample as possible group contaminants.
This method will obviously remove some true field
galaxies, as it does not account for the spatial positions of the possible group galaxies; 
however, this is a quick, simple method for removing galaxies that are
likely to live in groups.  We removed a total of 78 galaxies---or $\sim$25\%---of
our original field sample.

These observations of groups and the ``cleaned'' field sample (referred to simply as
the ``field'' from this point on) result in 
24\micron~measurements 
or upper limits for 232 group galaxies and 236 field galaxies from 
$0.3 \la z \la 0.55$, of which 79 group galaxies and 65 field galaxies are 
detected.  Our absolute detection limit (3$\sigma$) is 119$\mu$Jy, corresponding
to the 24\micron~observations with the highest background.  These observations 
result in a detection limit for the groups of
of L$_{TIR}$ $\sim$ 3.5 $\times$ 10$^{10}$ L$_{\odot}$ 
(SFR $\sim$ 3.3 M$_{\odot}$ yr$^{-1}$), where L$_{TIR}$ is the total IR luminosity
as defined by \citet{rieke09}.  However, only two groups (20 galaxies)
have detection limits at this level; $\sim$91\% of our group sample (and $\sim$93\% 
of the field sample) have 
24\micron~coverage to a lower detection limit:  L$_{TIR}$ $\sim$ 2.9 $\times$ 10$^{10}$ 
L$_{\odot}$ (SFR $\sim$ 2.7 M$_{\odot}$ yr$^{-1}$).  We use this lower value as our detection limit
for the rest of the paper.  At a typical redshift for our groups (z $\sim$ 0.43), L$_{TIR}^{*}$ 
corresponds to SFR $\sim$ 10 M$_{\odot}$ yr$^{-1}$ \citep{wiphu10}.
The typical optical surface density of the GEEC groups is $\sim$3
galaxies Mpc$^{-2}$ for galaxies with M$_{B_{J}}$ $<$ -20, though it ranges from $\sim$1 
to $\sim$6.5.  Cluster surface densities can range from these values in 
the outskirts to as high as several hundred galaxies Mpc$^{-2}$ in the dense cores
\citep{dressler80}.

\subsection{Uncertainties and Reliability}

We cross-correlated the sources detected at 24\micron~(almost 2000 objects over all fields)
with the GEEC spectroscopic catalog to within 3 arcsec; if
multiple optical sources were located within 3 arcsec of a 24\micron~source, the 
nearest optical source was used (only 20 instances of such multiple matches using the
entire 24\micron~and CNOC2 catalogs were recorded).
Any GEEC galaxy not matching this criterion was given a 3$\sigma$ upper limit.  
To estimate the 1$\sigma$ errors (and 3$\sigma$ upper limits for non-detected sources) 
in the 24\micron~flux densities, we
put down apertures in blank areas of the fields and took the standard deviation of the
nearest 20 background apertures to a given source.  

Because of the depth of our observations and the size of the MIPS 24\micron~PSF, it
is necessary to know the fraction of false detections---in other words, the fraction
of 24\micron~sources incorrectly attributed to an optical source in our catalog.
We therefore placed a random number of fake sources (up to 300 sources per pointing,
which at z = 0.55 equates a surface density of 4 Mpc$^{-2}$) on each image and
matched the 24\micron~detections to both the fake source catalog and our GEEC spectroscopic
catalog.  We use three different match radii---4, 3, and 2.5 arcsec---and found the mean fraction
of false detections to be 3.1\%, 1.8\%, and 1.3\%, respectively, for each match radius.
As expected, the number of
false detections per unit area is relatively constant and does not change
significantly for each match radius.  We could use any of the listed
match radii and not change the number of false matches substantially (i.e. from $\sim$3
to $\sim$7 for the group and field samples individually, out of $\sim$230 galaxies in 
each environment, and from $\sim$2 to $\sim$5 for the cleaned field). 
However, if we use too small a match radius ($\la$ 2.5 arcsec), we risk eliminating 
real matches. At our typical detection levels of $\sim$4--5$\sigma$, the rms positional
errors at 24\micron~are $\sim$1.5 arcsec.
Additionally, from examination of our HST ACS images (see the Appendix for
more information), we discovered that using a matching radius of 4 arcsec could
introduce more false detections than anticipated due to the crowded nature of our
group-dominated fields.  For these reasons, we used the 3 arcsec matching radius.
A 3 arcsec matching radius results in $\sim$4.2 incorrectly matched galaxies 
for the groups and $\sim$2.7 for the cleaned field for all galaxies.  
If we assume that all of the mis-matched galaxies are detected at 24\micron,
we would have upper limits that $\la$ 5.6\% and $\la$ 6.5\% of detected 
group and field galaxies, respectively, are incorrectly matched.

We also need to keep in mind that we are comparing group and field galaxies en masse, 
not individually.  False detections will affect the groups and field each the same
way, and so will not bias our results provided that the number of false detections remains low,
which we have already shown to be the case.  As such, the matching technique (and match 
length of 3 arcsec) is sufficient for our purposes.

\section{IR Properties of Sample Galaxies}

Extreme environments, such as the interior regions of clusters, clearly 
quench star formation in galaxies and reduce their IR outputs
(e.g., Bai et al. 2009).  Based on the ages of their dominant stellar populations and 
distributions of morphologies, galaxies in  
rich groups and clusters must have a similar evolution, at least at
low redshift \citep{bm10}.  
This suggests that galaxy transformation occurs in groups as well as in clusters.  
To test for such behavior, we compare
the fractional IR LF in groups with that of field galaxies 
at intermediate redshifts.

\subsection{Methodology}

To compare the IR output and SFR of group and field galaxies, we estimated the 
SFR and total IR luminosity (L$_{TIR}$) 
from the 24\micron~flux densities using the method described by \citet{rieke09},
who use {\it Spitzer} data to create a group of luminous 
star-forming galaxy templates more complete in the near- and mid-IR than 
previous templates.  

Our method for determining SFR and L$_{TIR}$ is only accurate for 
purely star-forming galaxies. 
Obscured active galactic nuclei (AGNs) can also emit a significant fraction of their light in the
mid-IR.  To identify possible AGN contaminants in our sample, we compared
our 24\micron-detected galaxies with {\it Chandra} detections.  
Four of the galaxies in the CNOC2 sample were identified as
AGN with clear X-ray detections, though all four were field galaxies outside 
of our redshift range (i.e. not in our GEEC sample).  
However, some AGN may be so buried that we are not able
to detect them at our current X-ray detection threshold.  More importantly, slightly
less than half of the groups targeted by MIPS have X-ray coverage, so there may 
still be some AGN contaminating our sample.  Given the few AGN detected with the
existing data and in previous studies of AGN activity in groups (Dwarakanath \& Nath 2006;
Silverman et al. 2009), we expect this number to be small.  

To correct for the different redshifts of our group and field galaxies,
we used the derived luminosity evolution of IR galaxies from \citet{lefloch05},
who show that IR galaxies as a whole evolve in luminosity as (1+z)$^{3.2}$
from z $\sim$ 0 to z $\sim$ 1.2.  We evolved our group and field galaxy IR 
luminosities to a fiducial redshift (z $= 0.5$) to remove redshift bias in
our sample.  This IR luminosity evolution is for all IR galaxies regardless
of environment; however, the \citet{lefloch05} sample (as for most field samples)
is expected to be $\sim$50\% groups, 
$\sim$50\% true field galaxies, making this model reasonable for our purposes.
This conclusion is supported by the lack of significant differences in the field
and group fractional luminosity functions (fLFs) as discussed below.

\subsection{Results for IR-active Galaxies}

\subsubsection{IR Luminosities and Star Formation Rates}

\begin{figure*}
\begin{center}
\centering
\includegraphics[angle=90,width=15cm]{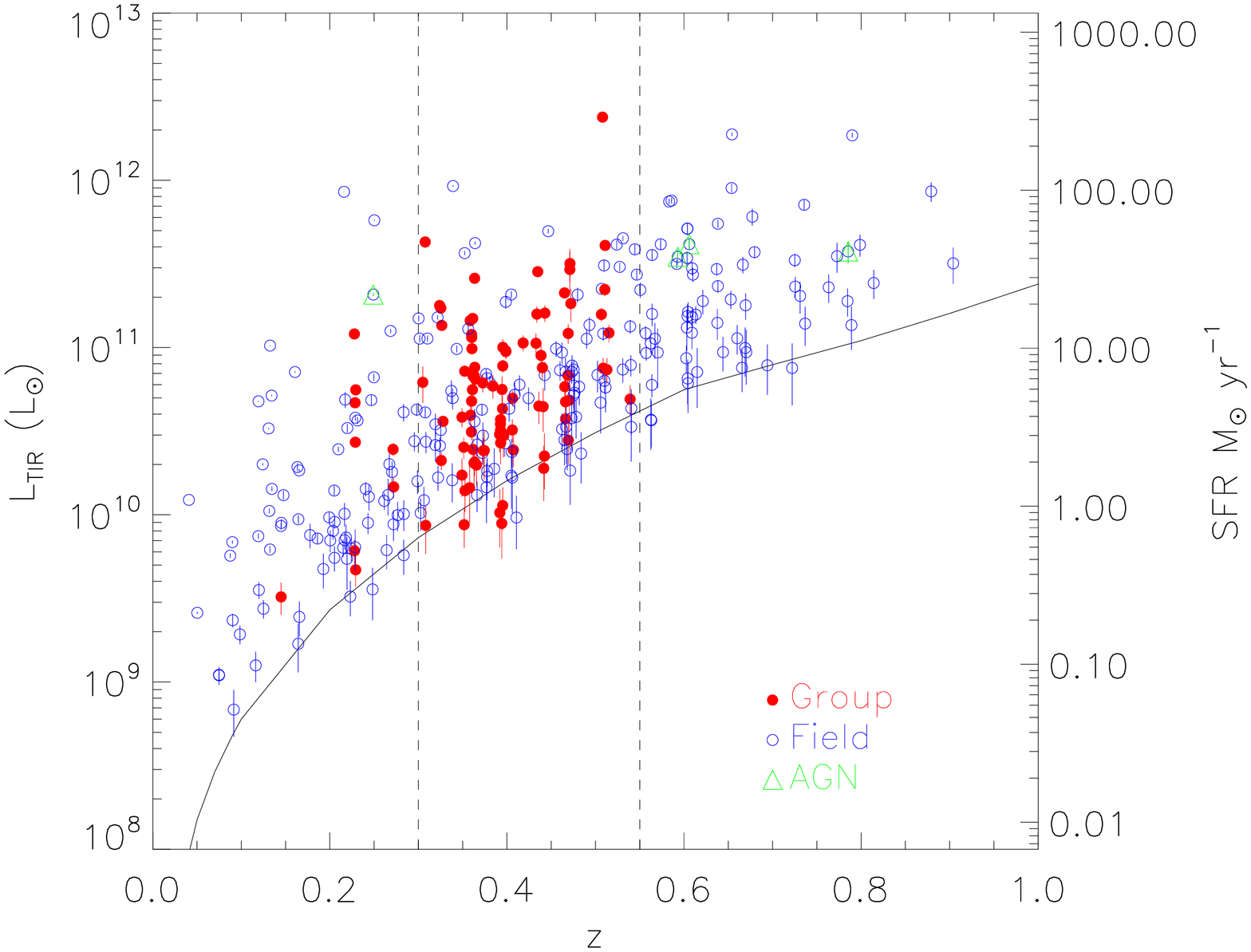}
\caption[]{L$_{TIR}$ and SFR versus redshift for all galaxies detected at 24\micron.  
The red filled circles are group members, the open blue circles are field galaxies, 
and the overplotted green triangles are X-ray-detected AGNs.  The dashed lines 
indicate the redshift range of our group and field sample (0.3 $\la$ z $\la$ 0.55).  
The solid line indicates the 
average 24\micron~3$\sigma$ detection limit for our observations.}
\label{fig:LTIRz}
\end{center}
\end{figure*}

In Figure \ref{fig:LTIRz} we present L$_{TIR}$ (left axis) and SFR (right axis)
with respect to redshift for all galaxies in the CNOC2 survey for which we
have MIPS 24\micron~detections.  Group galaxies are shown as solid red circles, while 
field galaxies are open blue circles and the X-ray identified AGN are overplotted as green
triangles.  The average detection limit for all of our 24\micron~fields 
is shown by the solid black curve; the redshift limits of our GEEC group and field sample 
are denoted by dashed lines ($0.3 \la z \la 0.55$).  The group and 
field IR-active populations do not appear significantly different.

\begin{figure*}
\begin{center}
\centering
\includegraphics[angle=90,width=15cm]{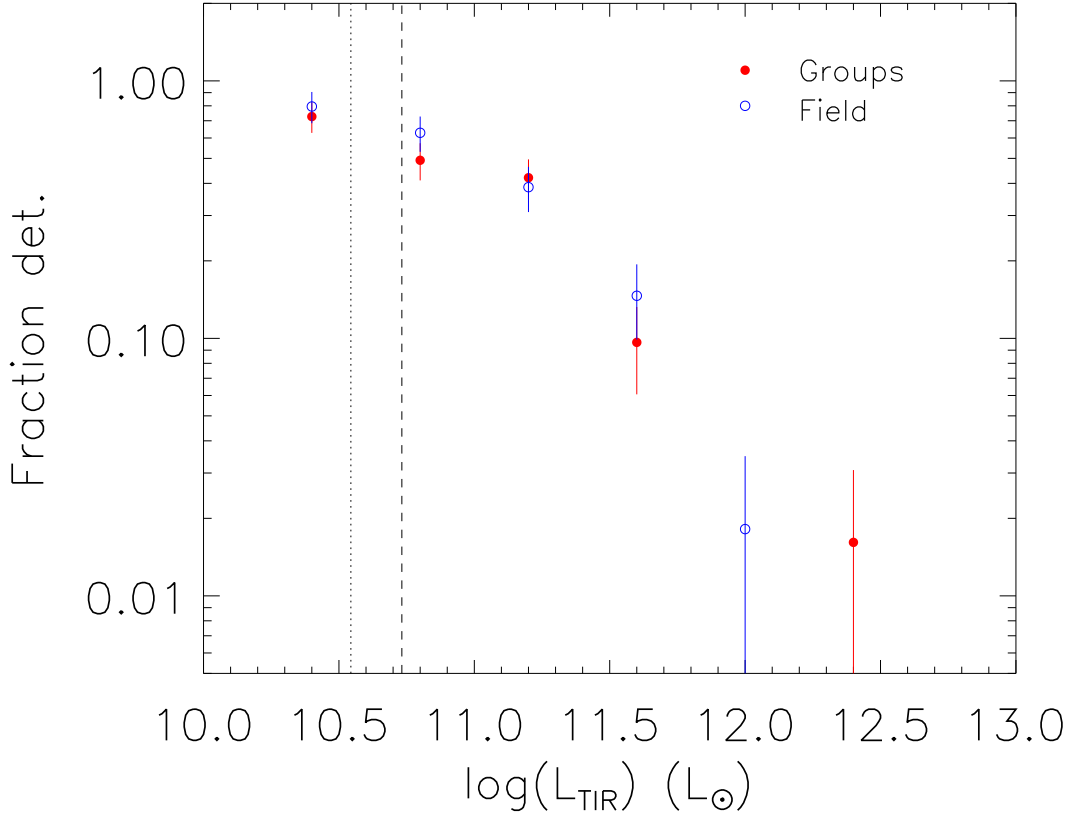}
\caption[]{L$_{TIR}$ histograms (what we refer to as fractional luminosity functions, 
fLFs) for the groups (red filled circles) and field galaxies
(blue open circles) with Poisson errors.
The galaxies included are those detected at 24\micron~in the redshift range 0.3 
$\la$ z $\la$ 0.55.  Each fLF is corrected for spectroscopic and 24\micron~completeness 
and has been normalized by the number of IR-detected galaxies in each environment.
The vertical dashed line is the detection limit for the field galaxies; the vertical
dotted line is the detection limit for all the group galaxies.
Up to log(L$_{TIR}$(L$_{\odot}$)) $\sim$ 12, the group and field fLFs are almost identical, given the
error bars.  The data for the group and field galaxies below the detection limits are
also in agreement, though there will be some bias as the group and field galaxies
are not equally complete.
Above log(L$_{TIR}$(L$_{\odot}$)) $\sim$ 12, we are limited by low number statistics, as there
are only a couple group and field galaxies at these luminosities.}
\label{fig:lf_LTIR}
\end{center}
\end{figure*}

Figure \ref{fig:lf_LTIR} gives us a clearer picture of the group and field 
populations in terms of L$_{TIR}$.  We plot histograms 
of the normalized distributions of L$_{TIR}$ (which we will refer to as fLFs from here on) 
for the group and field galaxies (red filled circles and
blue open circles, respectively).  
The dashed line indicates our 24\micron~detection limit for all field galaxies, 
while the dotted line is the detection limit for all group galaxies. 
The fractions in each bin (y-axis values) are normalized by the number of
24\micron-detected galaxies in each environment and corrected for spectroscopic incompleteness
as a function of magnitude. Weights were computed using the method outlined in 
Appendix A of Wilman et al. (2005a), except in this case we do not correct for any 
radial dependent selection, which tends to overweight galaxies on the outskirts of groups.

We have also corrected for completeness with respect to the 24\micron~data.
We calculated the 24\micron~incompleteness by estimating the SFR corresponding to 
the detection limits for all galaxies in each image.  We then
found the fraction of group and field galaxies (separately) with SFR detection limits 
below a given
SFR; the inverse of this fraction in each bin corresponds to our 24\micron~completeness 
weighting.  The completeness corrections for the points shown above both of
these limits are small:  factors of $\la$ 1.7 for both the group and field galaxies.

The group and field galaxy fLFs
are identical within the error bars up to log(L$_{TIR}$ (L$_{\odot}$)) $\sim$ 12.  
The last two bins contain few galaxies, so we cannot compare the bright
end with much certainty.
The measurements slightly below our detection limit should not be significantly
biased given the care to avoid bias in the sample selection \citep{wilman05} 
and that the vast
majority of the group and field galaxies have detection limits below the
indicated limit; therefore, we can say that these data also agree within the error bars
with some certainty.
Overall, the fLFs suggest that there is little difference between the groups
and the field (with respect to L$_{TIR}$).  Given that there are similar
numbers of spectroscopically-identified galaxies in the groups and field in this redshift
range, even the overall normalization will be the same.
This seems to indicate that the
group environment is not responsible for suppressing or enhancing star
formation amongst the actively star-forming population.    
To confirm the apparent similarity of the group and field samples, we performed a
two-sample Kolmogorov-Smirnov (K-S) test on the unbinned fLFs 
(the raw L$_{TIR}$ distributions) in 
both environments.   The K-S test indicated a
$\sim$99\% probability that the two distributions can come from the same parent 
sample.  Due to the coupling
between L$_{TIR}$ and SFR through L(24\micron) \citep{rieke09}, 
the results are the same if we plot SFR instead of L$_{TIR}$.


To better quantify the two LFs themselves, we fit both with a Schechter function, 
minimizing $\chi$$^{2}$.  Log(L$^{*}$ (L$_{\odot}$)) for the groups and field, respectively, are
11.9 $\pm$ 0.5 and 12.3 $\pm$ 0.3, a difference of $\sim$0.4 dex, though, given
the error bars, the two environments are similar.  \citet{tran09} find
the opposite trend:  their supergroup L$^{*}$ is $\sim$0.4 dex higher than
the field.  The faint-end slope
alpha does not differ significantly between the two environments either.
We find alpha to be 2.9 $\pm$ 1.1 and 4.3 $\pm$ 1.2 for the groups and field,
respectively.

\begin{figure*}
\begin{center}
\centering
\includegraphics[angle=90,width=15cm]{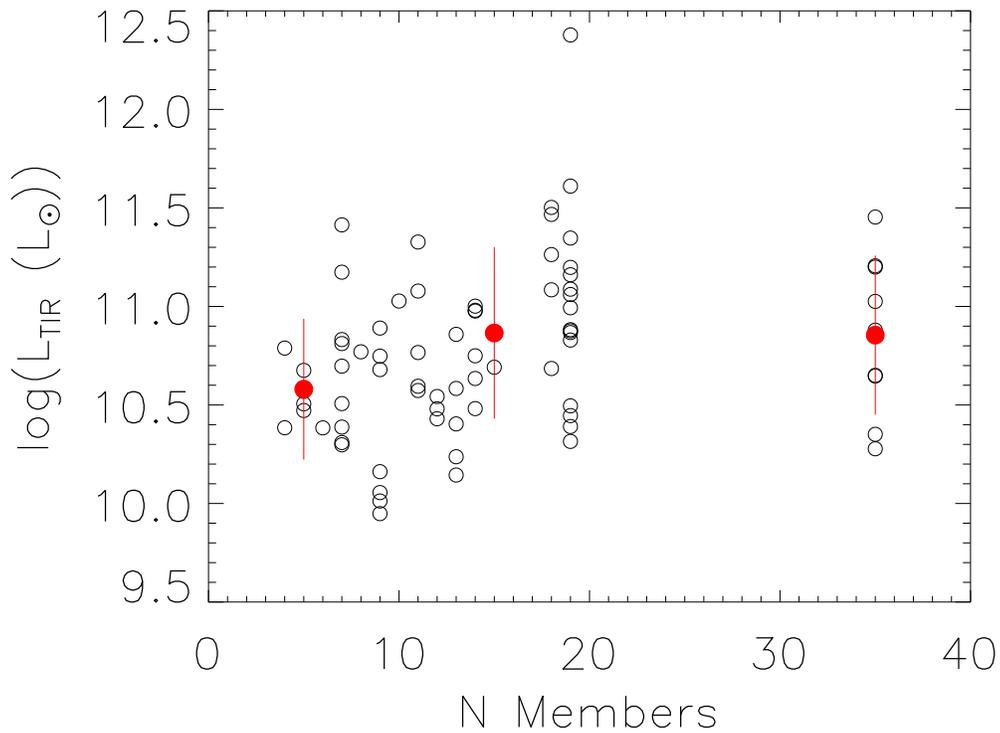}
\caption[]{L$_{TIR}$ of group galaxies detected at 
24\micron~(luminosity-evolved to z $\sim$ 0.5) 
versus the total number of members in each group brighter than 
M$_{B_{J}}$ $=$ -20.  The open black circles are individual galaxies; the red 
filled circles are the mean L$_{TIR}$ and 1$\sigma$ error bars for groups
in three different bins.  The total IR luminosity for individual
group galaxies does not depend on the size of the group.  In other words, we see no trend
in star-formation levels with group richness.}
\label{fig:NLTIR}
\end{center}
\end{figure*}

Figures \ref{fig:LTIRz} and \ref{fig:lf_LTIR}, and our statistical 
analysis of the results, suggest that the group environment does not 
substantially quench or enhance star formation at 
intermediate redshifts for the IR luminosities we are studying.  Could the
environmental effects be more prominent in the larger groups?
To test this possibility, we plot the number of group members brighter than M$_{B_{J}}$ $=$ -20
versus L$_{TIR}$ (luminosity-evolved to z $\sim$ 0.5) 
in Figure \ref{fig:NLTIR}.  The open circles are individual galaxies, and the
filled red circles show the mean L$_{TIR}$ in three bins for all detected group galaxies
with 1$\sigma$ error bars.  For reference, most groups have velocity dispersions
less than 500 km s$^{-1}$ while the largest group is $\sim$740 km 
s$^{-1}$(\footnote[4]{The group membership determination to
the limiting magnitude is complete to the $\sim$70\%--90\% level, so variations
in the completeness are not large enough to affect the basic result shown in the figure.
Ideally, we would compare L$_{TIR}$ directly with group mass determined from the 
velocity dispersions, but for values below $\sim$350 km s$^{-1}$ the velocity dispersions
can significantly underestimate the group mass \citep{nw87}.  Additionally,
for small groups, the orientation angle of the group on the sky may add uncertainty in
translating the velocity dispersion to the mass \citep{pt04}.  The velocity dispersions
for some of our groups have large formal errors, and a significant fraction of the 
groups may also not be virialized (e.g., Bai et al. 2009, Hou et al. 2009).}).
We see no trend in 24\micron-detected galaxies with respect to group size, and the
differences between the mean IR luminosity values is not significant.
Therefore, within the L$_{TIR}$ limit
of our sample, group size does not strongly affect the IR luminosity
of the member galaxies.  This is interesting given that interactions are thought
to be a major source of star-formation quenching in groups (given that higher
densities are needed for other quenching mechanisms like ram-pressure stripping and
strangulation), and the number
of galaxies that have experienced past interactions and mergers should be larger
in more massive groups, which have more members.

Our results are in agreement with many recent studies of clusters.
At similar redshifts to our groups (0.4 $\la$ z $\la$ 0.8), \citet{finn10}
find that clusters have IR LFs very similar to the field.  This
seems to be true with local clusters, as well (Bai et al. 2006, 2009), 
indicating that IR-active galaxies in clusters have been
recently-accreted and have not yet had time to become affected by the
more dense environment of the cluster.  \citet{cortese08}
find that the UV LF of the Coma cluster is indistinguishable from the 
field; however, they argue that this is due more to color selection effects than
environmental processes.  We believe that the GEEC R$_{C}$ selection does
not cause significant selection effects, but further study would be needed to 
determine how this R-band selection might affect our results.

\subsubsection{Star Formation with Respect to Mass}
 
While the total IR luminosities and SFRs of the group galaxies do not appear 
different from the field, there could be trends with stellar mass.  
The most massive galaxies in dense regions of the local universe, such as cluster cores,
are old ellipticals, whereas the most massive field galaxies are blue spirals.  
What about groups at intermediate redshifts?  

The stellar masses used here are presented in \citet{mcgee}, but
briefly, we use spectral energy distribution (SED) fitting of all
available photometry. The details of the photometry were presented in
Balogh et al. (2009) but typically involved K, i, r, g, u, NUV and
FUV. We compare this observed photometry to a large grid of model SEDs
constructed using the Bruzual \& Charlot (2003) stellar population
synthesis code and assuming a Chabrier initial mass function. We
follow Salim et al (2007) in creating a grid of models that uniformly
samples the allowed parameters of formation time, galaxy metallicity,
and the two components of the Charlot \& Fall (2000) dust model. The
star formation history is modeled as an exponentially declining base
rate with bursts of star formation randomly distributed in time, 
which vary in duration and relative strength. We produce model
magnitudes by convolving these model SEDs with the observed
photometric bandpasses at nine redshifts between 0.25 and 0.6. Finally,
we minimize the $\chi^2$ while summing over all the models and taking
account of the observed uncertainty on each point. By comparing to
other estimates of stellar mass we estimate that 1$\sigma$ uncertainties
are on the order of 0.15 dex.

\begin{figure*}
\begin{center}
\centering
\includegraphics[angle=90,width=15cm]{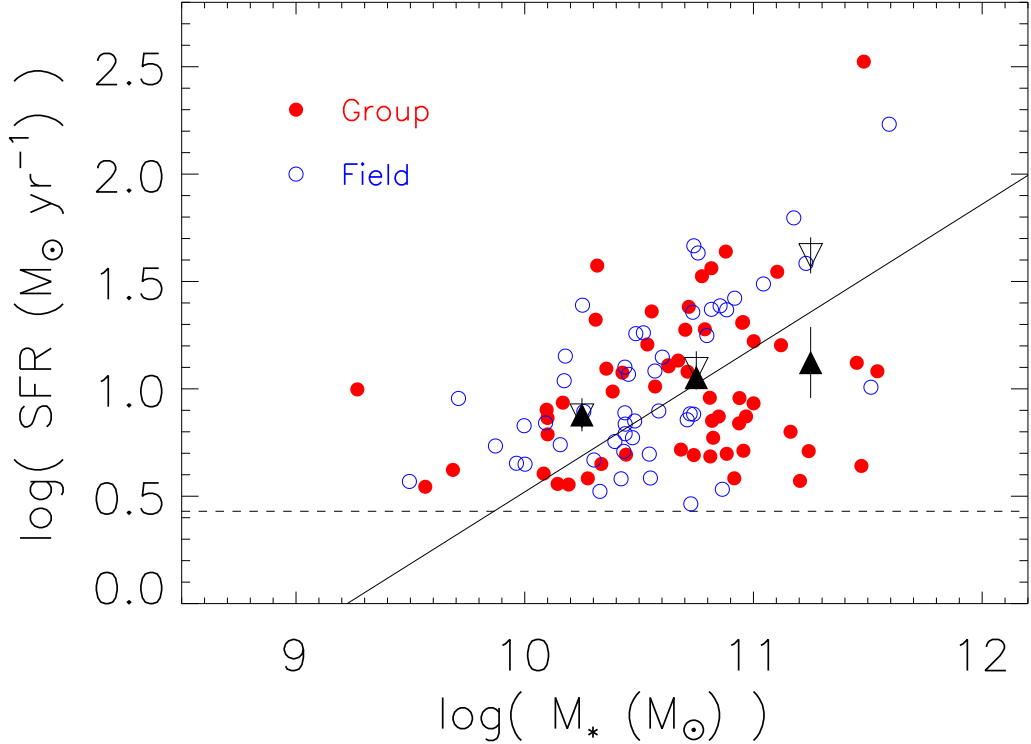}
\caption[]{Stellar mass versus SFR for IR-detected galaxies in groups (red filled 
circles) and the field (blue open circles).  We only show galaxies above our 
24\micron~detection limit, which is shown by the dashed line.  The black filled 
triangles (black open triangles) are the mean SFR for the groups (field) in three 
mass bins above log(M$_{*}$(M$_{\odot}$)) $=$ 10.  The field and group galaxies 
have different distributions.
The solid line represents the trend found by \citet{noeske07} for 24\micron~identified
``field'' galaxies at 0.2 $\la$ z $\la$ 0.7 (plotted here at an arbitrary normalization).  
Because \citet{noeske07} did not distinguish between group and field galaxies, their trend is likely
a combination of the two environments.}
\label{fig:StMass}
\end{center}
\end{figure*}
\begin{figure*}
\begin{center}
\centering
\includegraphics[angle=90,width=15cm]{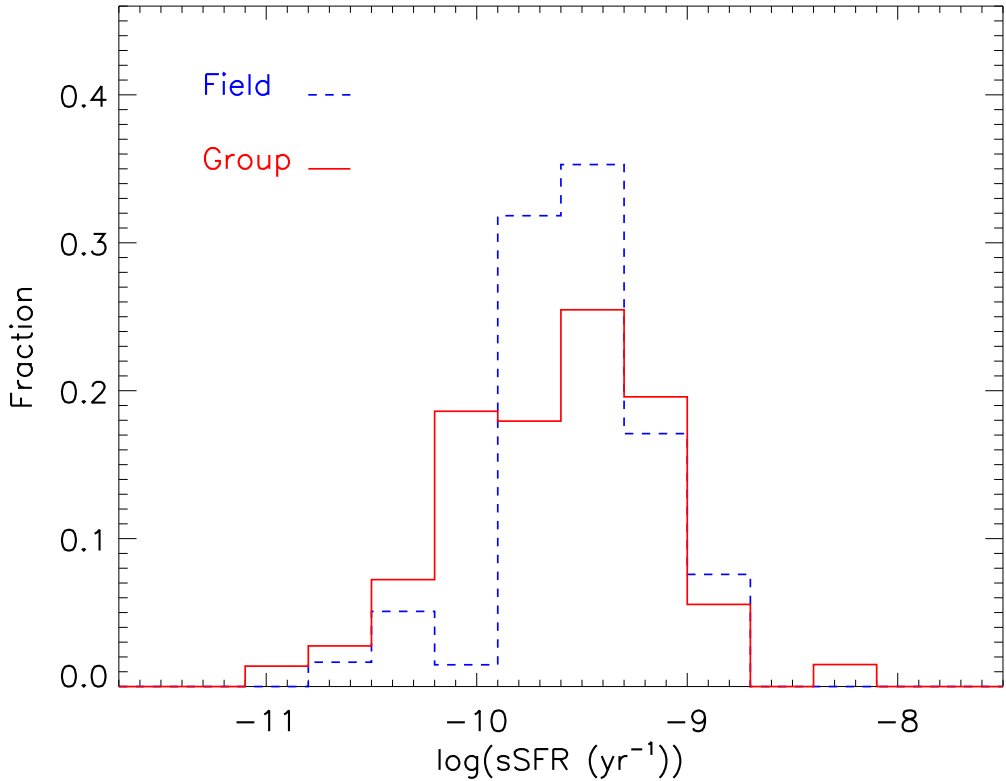}
\caption[]{Histograms of specific star formation rate (SFR M$_{*}^{-1}$) for 
IR-detected galaxies (SFR $\ga$ 2.7) in groups (red solid line) and the 
field (blue dashed line), all corrected for spectroscopic and 24\micron~incompleteness.  
Each histogram has been normalized so its total value is 1.  The groups have more 
galaxies at higher masses than their field counterparts, resulting
in lower specific SFRs for the groups.  This difference is not highly 
significant, however:  a two-dimensional KS test results in a 27\% probability
that these two data sets come from the same distribution.}
\label{fig:sSFR}
\end{center}
\end{figure*}
\begin{figure*}
\begin{center}
\centering
\includegraphics[scale=0.6,width=15cm]{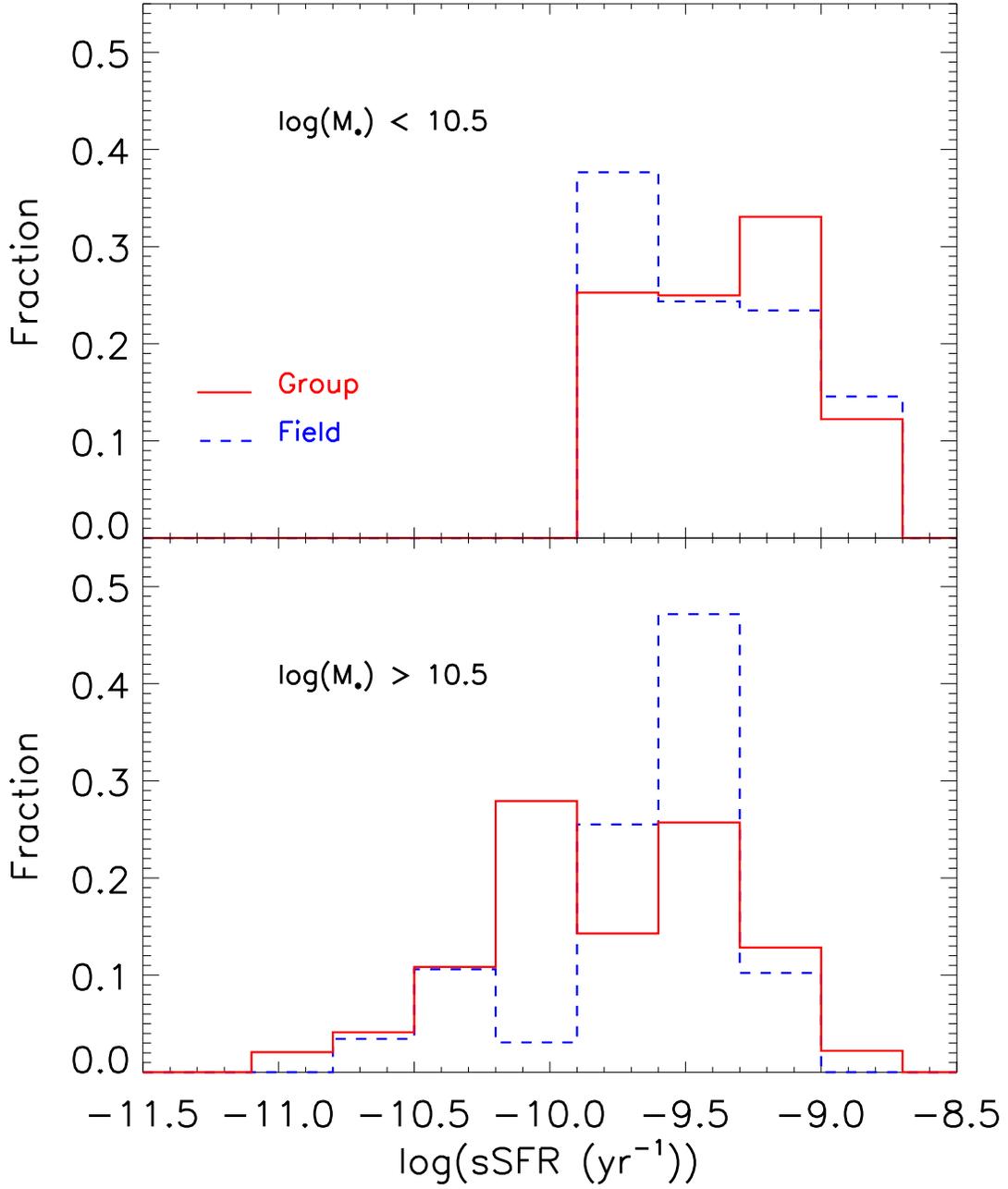}
\caption[]{Histograms of specific SFR for the field and group galaxies with
log(M$_{*}$ (M$_{\odot}$)) $<$ 10.5 (top plot) and $>$ 10.5 (bottom
plot), with the same normalization and completeness corrections as the previous
figure.  While the ranges of sSFRs for both environments are similar,
we can see a weak trend whereby massive field galaxies have higher sSFRs
than the groups.  This same trend was evident in Figure \ref{fig:StMass},
where we can see that the groups tend to have more massive galaxies with
lower SFRs than the field.
A two-sample K--S test comparing the high-mass group and field
galaxies results in a $\sim$3\% probability that the two populations come from 
the same distribution, and the low-mass group and field galaxies have an 
89\% probability of coming from the same distribution.}
\label{fig:SFR_mass}
\end{center}
\end{figure*}

We investigate SFR with respect to stellar mass
in Figures \ref{fig:StMass} through \ref{fig:SFR_mass}.  
Figure \ref{fig:StMass}
shows SFR plotted with respect to stellar mass for the group (red filled circles)
and field (blue open circles) galaxies.  Only galaxies above our 24\micron~detection 
limit (dashed line) are shown.  Because our sample is only unbiased for
log(M$_{*}$(M$_{\odot}$)) $\ga$ 10, we average the SFRs for
each environment in three mass bins above this limit.  These averages are
plotted as black filled triangles for the groups and black open triangles for the field.
Unlike the previous figures, the field has a different distribution
than the groups.  \citet{noeske07} compare SFR and stellar mass
for 24\micron-identified ``field'' galaxies in a similar redshift range 
(0.2 $\la$ z $\la$ 0.7) and find a linear relation between SFR and stellar
mass with a slope of $\sim$0.67.  They did not distinguish
between group and field galaxies, however, so their ``field'' is a 
combination of the two environments.  We compared our galaxies with theirs
by plotting this relation as a solid, black line (with an arbitrary normalization).
Our group and field galaxies, combined, seem to echo the \citet{noeske07} relation,
with the suggestion that galaxies in the field obey a steeper mass--SFR relationship 
than those in groups (mainly driven by the lack of high-mass galaxies with SFR just
above our detection limit in the field, a region populated by group galaxies).

In Figure \ref{fig:sSFR}, 
we plot a ``specific SFR function'' (SFR M$_{*}^{-1}$)
in the same manner as Figure \ref{fig:lf_LTIR}, though only including galaxies
with SFR $\ga$ 2.7 M$_{\odot}$ yr$^{-1}$.  Each histogram has been normalized
so that its total is 1.  Differences between the group and the 
field are suggested; a two-sample K-S test reveals that the two distributions
have only a 27\% probability that they are drawn from the same parent sample.  
The groups have more IR-active galaxies with less star formation per stellar mass
than the field galaxies, as expected given the overall higher masses of the
group galaxies.  Put another way, for the IR-detected galaxies in the
groups, the on-going star formation makes a smaller relative contribution
to the stellar mass than for galaxies in more isolated environments.
Thus, the average timescale for growth of the stellar mass ($M_* / \dot{M}_*$) is 
currently smaller in the field than in group galaxies by a factor of $\sim$3. In the
past, however, it is possible that this timescale was shorter in the groups given that their
galaxies have shifted to higher masses by z $\sim$ 0.5.

Figure \ref{fig:SFR_mass} compares stellar mass and SFR
in a slightly different way.  The top plot shows the fraction of 
24\micron-detected field and
group galaxies as a function of specific SFR (sSFR) 
with log(M$_{*}$(M$_{\odot}$)) $<$ 10.5, labeled as
blue dashed and red solid lines, respectively.  The lower panel is the same plot
except for galaxies with log(M$_{*}$(M$_{\odot}$)) $>$ 10.5.  The histograms
in this figure have been normalized in the same manner as Figure \ref{fig:sSFR}. 
We see that group and field galaxies have similar ranges of sSFRs in either
mass bin, but the higher-mass galaxies in the field tend to form stars at higher
rates (higher-mass group galaxies have lower relative SFRs).  
A two-sample K-S test results in a $\sim$3\% probability
that the low-mass (log(M$_{*}$(M$_{\odot}$)) $<$ 10.5) group and field galaxies come from
the same parent sample and a $\sim$89\% probability for the high-mass
group and field galaxies (log(M$_{*}$(M$_{\odot}$)) $>$ 10.5). 
While neither of these cases indicate a 3$\sigma$ result, it makes sense given
Figure \ref{fig:StMass}, which shows a trend of higher-mass group galaxies
having significantly lower SFRs as compared with the field.  The subtle trends seen in Figures
\ref{fig:sSFR} and \ref{fig:SFR_mass} may be showing the beginning
stages of suppression of star formation in the groups.

\begin{figure*}
\begin{center}
\centering
\includegraphics[angle=90,scale=0.6,width=15cm]{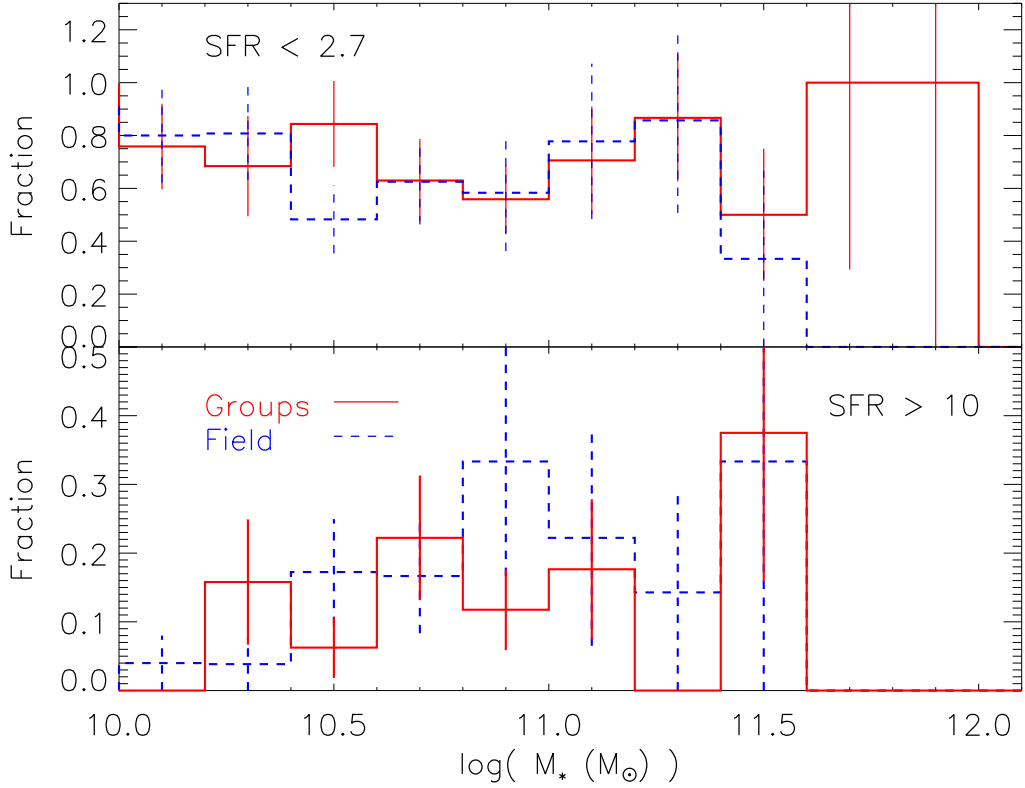}
\caption[]{Fractions of low- and high-activity galaxies with respect to environment.  The top
plot shows the fraction of galaxies with SFR $<$ 2.7 M$_{\odot}$ yr$^{-1}$ normalized by the
total number of galaxies in each environment, respectively, per mass bin.  For stellar
masses below $\sim$11.5, the groups and the field are nearly identical; above this limit,
however, the groups have a few massive, low-activity galaxies while the field has none.  
(This is not significant, as there are only three galaxies in the two highest mass bins 
for the groups.)  The bottom plot shows the fractions
of galaxies with SFR $>$ 10 M$_{\odot}$ yr$^{-1}$ for each environment, respectively.  
We see a stronger difference between the group and field galaxies here than in the 
top plot:  the groups have 
galaxies forming stars at this high rate at a variety of masses, while the field galaxies
peak at log(M$_{*}$(M$_{\odot}$)) $\sim$ 11.  
This may be another indication of a stronger mass-SFR
relation in the field than the group galaxies, though the significance is small.}
\label{fig:morph_hist}
\end{center}
\end{figure*}

In Figure \ref{fig:morph_hist}, we split our group and field galaxies in terms
of their SFRs with respect to stellar mass.  The top plot shows the fraction
of what we call ``low-activity'' galaxies (SFR $<$ 2.7 M$_{\odot}$ yr$^{-1}$)
normalized by the total number of galaxies per mass bin in the groups (red solid
line) and field (blue dashed line).
The two highest mass bins in the groups
house a total of three galaxies, all of which are in groups with 10 or fewer
members, so any trend in the number of low-activity, high-mass group galaxies
does not appear to be significant.  As a result, the group and field galaxies
are similar in terms of the fraction of low-activity galaxies 
for a given stellar mass.
The bottom plot shows the fraction of galaxies in each environment with SFR $>$
10 M$_{\odot}$ yr$^{-1}$ with the same normalization as the top plot.
We do potentially see a stronger difference between the group and field galaxies at these
higher SFRs:  the groups have galaxies forming stars at this high rate
at a variety of masses, while the field galaxies peak at 
log(M$_{*}$(M$_{\odot}$)) $\sim$ 11.
This may be another indication of a stronger mass--SFR relation in the 
field than in the groups, though the differences between the group and
field galaxies here are of low significance.

\begin{figure*}
\begin{center}
\centering
\includegraphics[angle=90,scale=0.6,width=15cm]{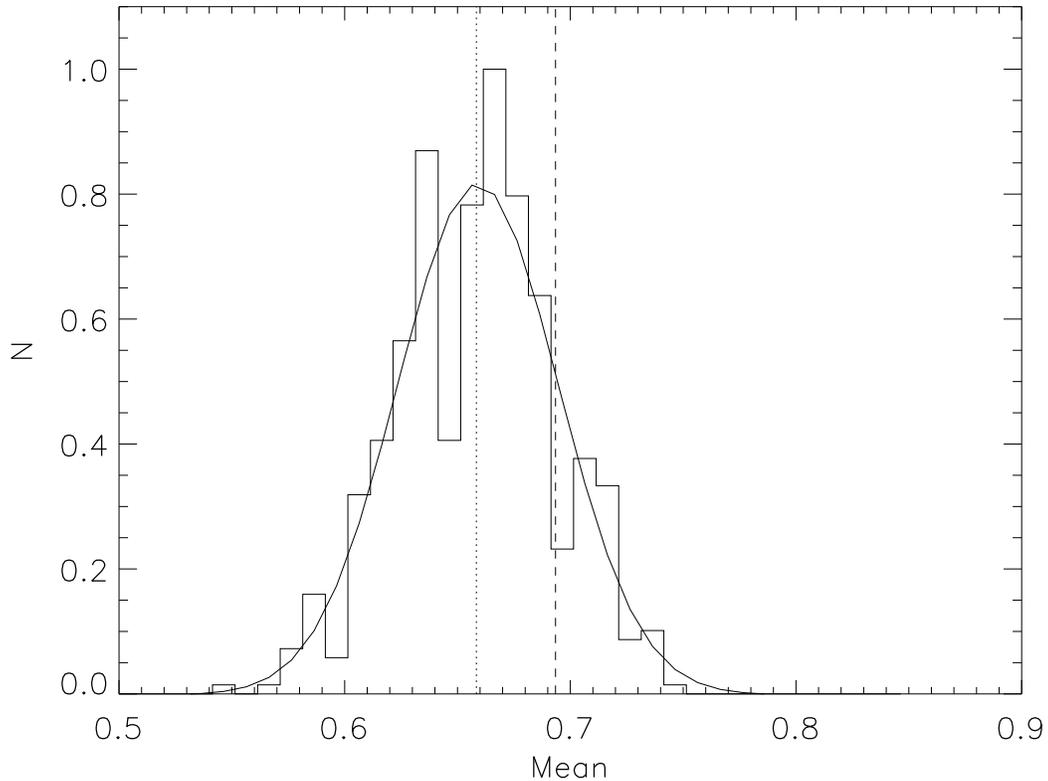}
\caption[]{Quantitative assessment of the top plot in Figure \ref{fig:morph_hist} (see the text
for a more specific explanation).  We created fake group samples from the field and calculated
the average fraction of low-activity galaxies from 10 $\la$ 
log(M$_{*}$(M$_{\odot}$)) $\la$ 11.6.  The histogram
shows the distribution of these averages for 500 fake group samples.  The dashed (dotted) 
line is the average fraction of low-activity galaxies in the real groups (field).  
The fake groups have 
consistently low averages as compared with the real groups, but this difference is only significant
at a 1$\sigma$ level, as shown by the Gaussian fit to the distribution (solid curve).}
\label{fig:mean}
\end{center}
\end{figure*}

We performed a Monte Carlo analysis of the data in the top plot of 
Figure \ref{fig:morph_hist} to estimate whether the overall fraction of low-activity
galaxies, at constant stellar mass, is significantly higher for group galaxies.
We found the total number of group
galaxies per mass bin and randomly selected the same number of galaxies per
mass bin from the field, making a fake group sample.  
We then made the same plot as the top part
of Figure \ref{fig:morph_hist} using the fake group galaxies:  the fraction of fake
low-activity galaxies normalized by the total number of galaxies per mass bin.
We repeated this 500 times, each time calculating the mean fraction of low-activity
galaxies from 10 $\la$ log(M$_{*}$(M$_{\odot}$)) $\la$ 11.6.  (Higher masses were not possible
given the lack of field galaxies in the highest mass bins.)  Because the fraction 
of low-activity galaxies in the groups and field in this mass range is fairly constant with mass,
the average is an accurate way of comparing the fake and real group galaxies.
This distribution of fake group averages is plotted as a histogram in Figure 
\ref{fig:mean}; it is fit well by a Gaussian (solid curve).
The mean of the real groups (real field) is shown as a dashed (dotted) line.  The fake groups have
consistently low averages as compared with the real groups, though this difference
is only significant to a 1$\sigma$ level.  Interestingly, the resampled field (matched
in mass to the group population) has a mean low-activity fraction equivalent to the
non-resampled field, indicating that the dependence of low-activity fraction on mass
within this range is negligible (also evidenced by the lack of a trend in Figure 
\ref{fig:morph_hist}).

Another way to describe these results is that, if we use the field mass-SFR relation 
to determine the expected IR luminosity distributions in groups, we would conclude 
that the groups are slightly under-luminous because of their higher proportion
of high-mass galaxies.
\citet{marcillac08} study the environment of 0.7 $\la$ z $\la$ 1.0 LIRGs and
ultraluminous IR galaxies (ULIRGs), and they find that, 
at similar masses, 32\% of all the galaxies 
in their sample reside in groups, and 32\% of their LIRGs and ULIRGs also reside
in groups.  Where we find a small and barely significant difference, their 
study indicates none.  That is, 
this reinforces our conclusion that the group environment does not suppress or 
enhance star formation in the galaxies as a whole.  Though we do see indications
of suppression in the groups when we compare sSFRs, 
the difference is subtle and might depend on 
the low-mass regime where we are incomplete (galaxies with lower SFRs than
we were able to detect with MIPS), or other variables.

\subsubsection{Morphologies}

Another parameter subject to transformation in dense environments is galaxy morphology.
Is there any difference in morphology types with respect to star formation
 between the group and field galaxies?  
\citet{wilman09} report on the visual morphologies of the GEEC groups and field
galaxies using high-resolution ACS data.  They find that the two environments 
tend to harbor different types of galaxies:  S0 galaxies are more prevalent in groups than
in the field at a fixed luminosity, indicating that suppression of star formation
and bulge growth have been more common in the group environment.  
Overall, for galaxies brighter than M$_{r_{0}}$ $=$ -21,
they find that the groups have about 1.5 times 
the number of E/S0 galaxies as the field \citep{wilman09}.

We compared the visual morphologies with our IR SFRs using the \citet{wilman09}
classifications.  Thus, ``early-type spirals'' (eSp) are galaxies
classified as Sa through Sbc (including barred spirals), and 
``late-type spirals'' (lSp) are classified as Sc through Sm (including barred spirals).
The fractions of IR-detected galaxies with optical classifications (both groups and
the field)
are as follows:  13\% of ellipticals, 10\% of S0s, 60\% of
eSps, 33\% of lSp, 15\% of irregulars, 17\% of mergers, 
and 61\% of galaxies identified as ``peculiar.''  
It is perhaps a surprise that eSps are twice as likely to be detected at 
24\micron~as lSps, but this difference may arise because of the lower masses of 
the later galaxy types. 

\begin{figure*}
\begin{center}
\centering
\includegraphics[scale=0.6]{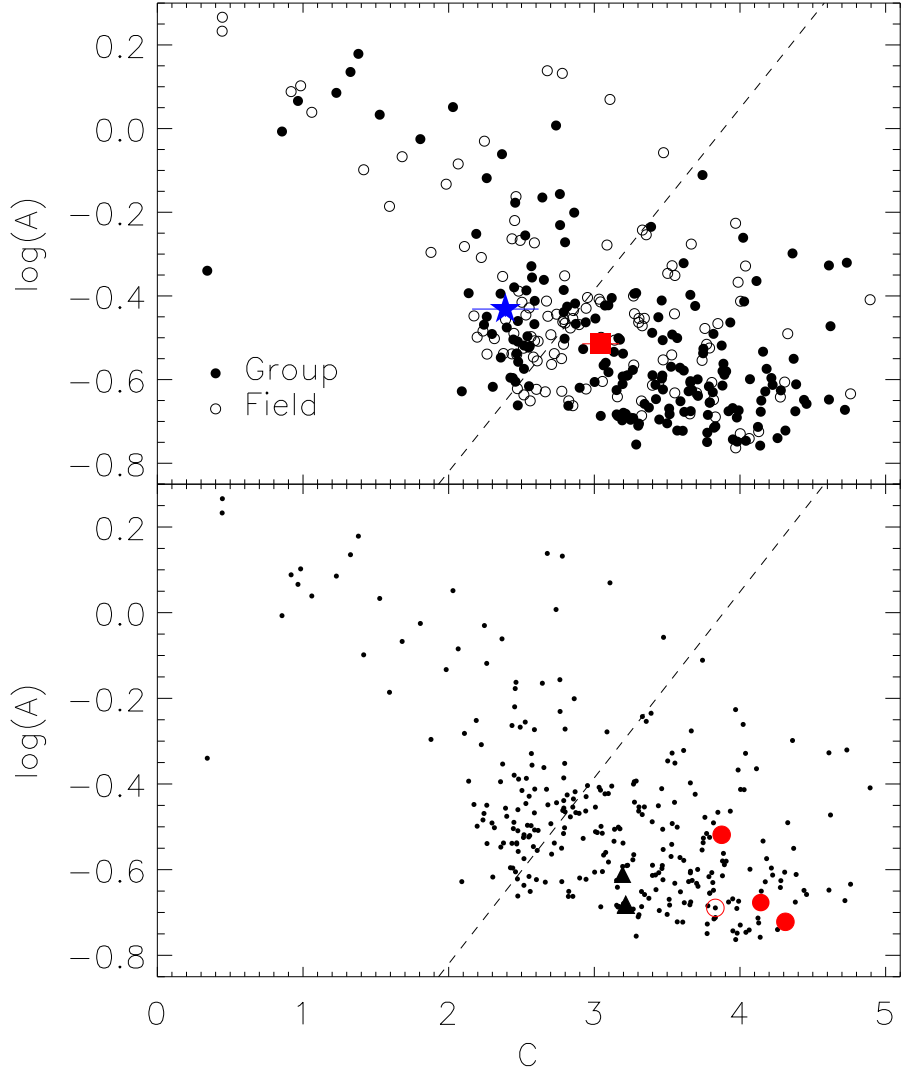}
\caption[]{Top:  asymmetry versus concentration for group (black filled circles) 
and field (black open circles) galaxies, calculated in the same manner as \citet{shi09}.  
These two parameters allow quantitative assessments of galaxy morphology (as opposed 
to qualitative visual classifications that can be biased by surface brightness dimming).  
Average values are 
shown as a red square (groups) and blue star (field), with error bars.  
The dashed line in both plots is an arbitrary division between the majority of the
group and field galaxies, where most of the galaxies below this line are visual early-types
and most of the galaxies above this line are late-types.
This plot confirms the trend in the visual morphologies:  the groups have
a higher fraction of early-type galaxies than the field, as shown by the
low asymmetry and high concentration of the groups.
Bottom:  asymmetry versus concentration for group and field galaxies (small black
dots), as with
the top figure.   Our six IR-active E/S0 galaxies are shown 
as red circles and black triangles, respectively, with group early-types having 
filled symbols and field galaxies having open symbols.  All six E/S0 galaxies
fall in the area of the plot populated by early-type galaxies, showing that 
these IR-active galaxies are indeed early-types.}
\label{fig:ac}
\end{center}
\end{figure*}

These conclusions can be tested with quantitative morphology
metrics, such as the Concentration, Asymmetry, Clumpiness method 
(Abraham et al. 1994, 1996; Conselice et al. 2000, 2003; see also McGee et al.
2008 for other methods).  
At z $\sim$ 0.5, surface brightness dimming might become a problem when identifying
spirals, particularly late-types, as the fainter spiral features become
indistiguishable from the background.
\citet{shi09} have shown how to correct concentration and asymmetry (referred to
as CA from now on) for the effects of surface brightness 
dimming to provide unbiased metrics at and
above the redshifts of our groups.  We have used their methodology to calculate CA
for our sample of field and group galaxies, as shown in
Figure \ref{fig:ac} (small black dots).  In the top plot, the 
average values (and error bars) for the groups and field galaxies 
are given by the red square and blue star, respectively.  
The dashed line is an arbitrary division roughly separating early- and late-type galaxies
that we discuss below.  Consistent with 
\citet{wilman09}, our results indicate a tendency for group galaxies to have, on average, a 
higher concentration and lower asymmetry (indicative of E and S0 galaxies) than in the
field.  

Of 144 galaxies detected at 24\micron, seven are Es and four are S0s.
Three of the galaxies are classified as ``peculiar,'' which means they have 
been visually identified as either having an interacting neighbor or a morphology 
slightly disturbed from the given classification, and one galaxy is listed as 
an S0/Sa (most likely an S0 but also has Sa qualities).  Still, it is  
unusual to find any E or S0 galaxies with SFRs at these levels.
To be certain the early-type galaxy detections are robust, we investigated
other possibilities for the IR emission.

The first issue concerns our source matching:  is it possible that
there are nearby (projected) neighbors that are being mis-matched with the
E/S0 galaxies?  As discussed in the Appendix, inspection of the ACS images 
shows that a few of the early-type galaxies have close neighbors, making
identifying which galaxy in the field of view is responsible for the IR
emission difficult.  To be conservative, we rejected any early-type galaxies 
where the IR emission could be coming from another object. 
After removing these ambiguous E/S0 detections, we were
left with four ellipticals and two S0s firmly detected at 24\micron.
We also double-checked our matching of the 24\micron~positions with
the IRAC 3.6\micron~coordinates.  Four of the six E/S0 galaxies have
IRAC coverage, and all of them match the IRAC coordinates within
1.5 arcsec or less. Because we have imposed very stringent requirements to 
claim a detection, six detected E/S0 galaxies represents a lower limit.

As a double--check on the morphologies, we again plot asymmetry and 
concentration for all galaxies with CA values (small, black circles)
and the six IR-active early-type galaxies (red circles for ellipticals
and black triangles for S0s, with filled points indicating a
group galaxy and empty points indicating a field galaxy) at the 
bottom of Figure \ref{fig:ac}.  All six galaxies fall in the 
``early-type'' area of the plot, and they also all fall well under
our early-/late-type dashed line.  It appears that these six
E/S0 galaxies are indeed early-types with bright 24\micron~emission.

\begin{figure*}
\begin{center}
\centering
\includegraphics[width=15cm]{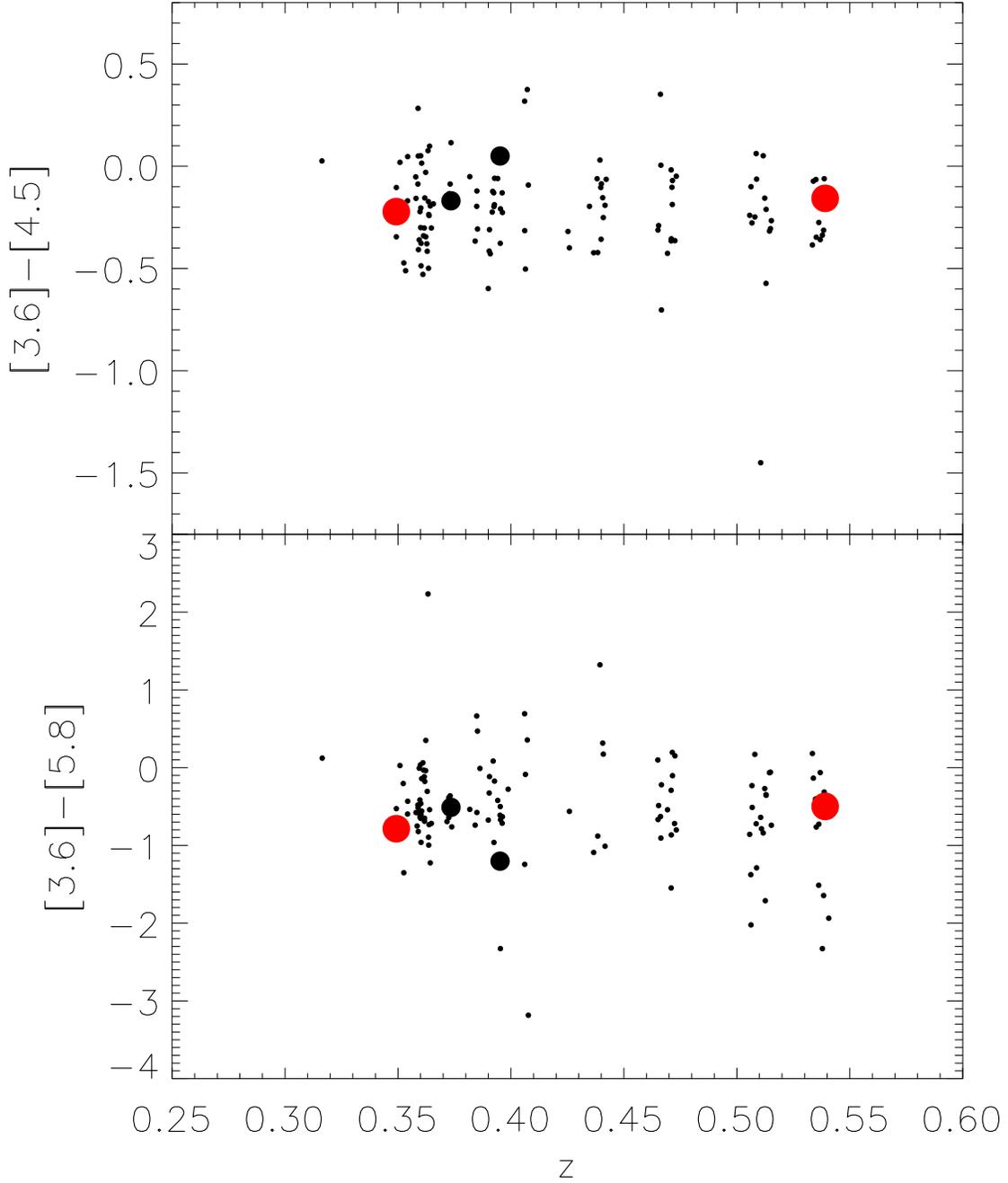}
\caption[]{IRAC colors [3.6]--[4.5] (top) and [3.6]--[5.8] (bottom) versus z for all galaxies
in our sample with IRAC coverage in those bands (small black circles).  Also included
are four of the six IR-active early-type galaxies that have IRAC coverage.  The S0 galaxies
are denoted by medium-sized black circles while the ellipticals are denoted by large red circles.
Galaxies dominated in these bands by warm dust from an AGN will have highly negative values
of [3.6]--[4.5] and [3.6]--[5.8] due to the SED brightening through the IRAC bands.
The four E/S0 galaxies with IRAC coverage lie with the rest of the ``normal''
galaxies, showing that the 24\micron~emission from these galaxies is dominated
by star formation and not AGN activity.}
\label{fig:irac}
\end{center}
\end{figure*}

There is one other culprit that could be masquerading as star formation, 
however:  AGNs.
Of the six early-type galaxies with confirmed IR emission, only two have 
X-ray coverage, though neither of them are detected down
to L$_{X}$ $\sim$ 10$^{41}$ erg s$^{-1}$ \citep{mulchaey}. 
As an additional test for AGNs, we look to the IRAC data.
The intersection of the stellar light and the warm/cool dust components
produces a dip in the SED at rest-frame $\sim$5\micron~for normal star-forming galaxies.
Buried AGNs heat the dust to higher temperatures, which ``fills in'' this
dip. If our IR-detected early-type galaxies have similar colors
at IRAC wavelengths to the typical galaxies in our sample, 
then AGNs are probably not contributing significantly to the mid-IR emission.  
In Figure \ref{fig:irac} we plot IRAC [3.6]--[4.5] and [3.6]--[5.8]
versus redshift for all galaxies with IRAC coverage in our sample.
Most of the galaxies fall into a narrow color range in both plots, 
indicating that we are probing the Rayleigh--Jeans tail of the stellar bump. 
If the 24\micron~emission were dominated by an embedded AGN, 
these colors would be more positive due to the warm dust filling in 
the ``dip'' in the SED, causing the SED to brighten as we move through
the IRAC bands.  None of the early-type galaxies detected at 
24\micron~(with IRAC coverage) show the signature of an AGN in the IRAC colors.

Two of our six IR-bright early-type galaxies do not have X-ray or complete 
IRAC coverage.  We have enough SED coverage of one of these two to make out
some of the stellar bump and the increase in dust emission in the IR, which
seems to indicate star formation instead of an AGN, but it is difficult to say 
with certainty.  The other source has too few photometric data
points to make any solid conclusion; however, given the lack of AGNs we 
have found so far in our entire sample, it seems that AGNs are a rarity.
While we cannot be certain that these two galaxies do not host AGNs, it seems
unlikely.

\begin{figure*}
\begin{center}
\centering
\includegraphics[width=15cm]{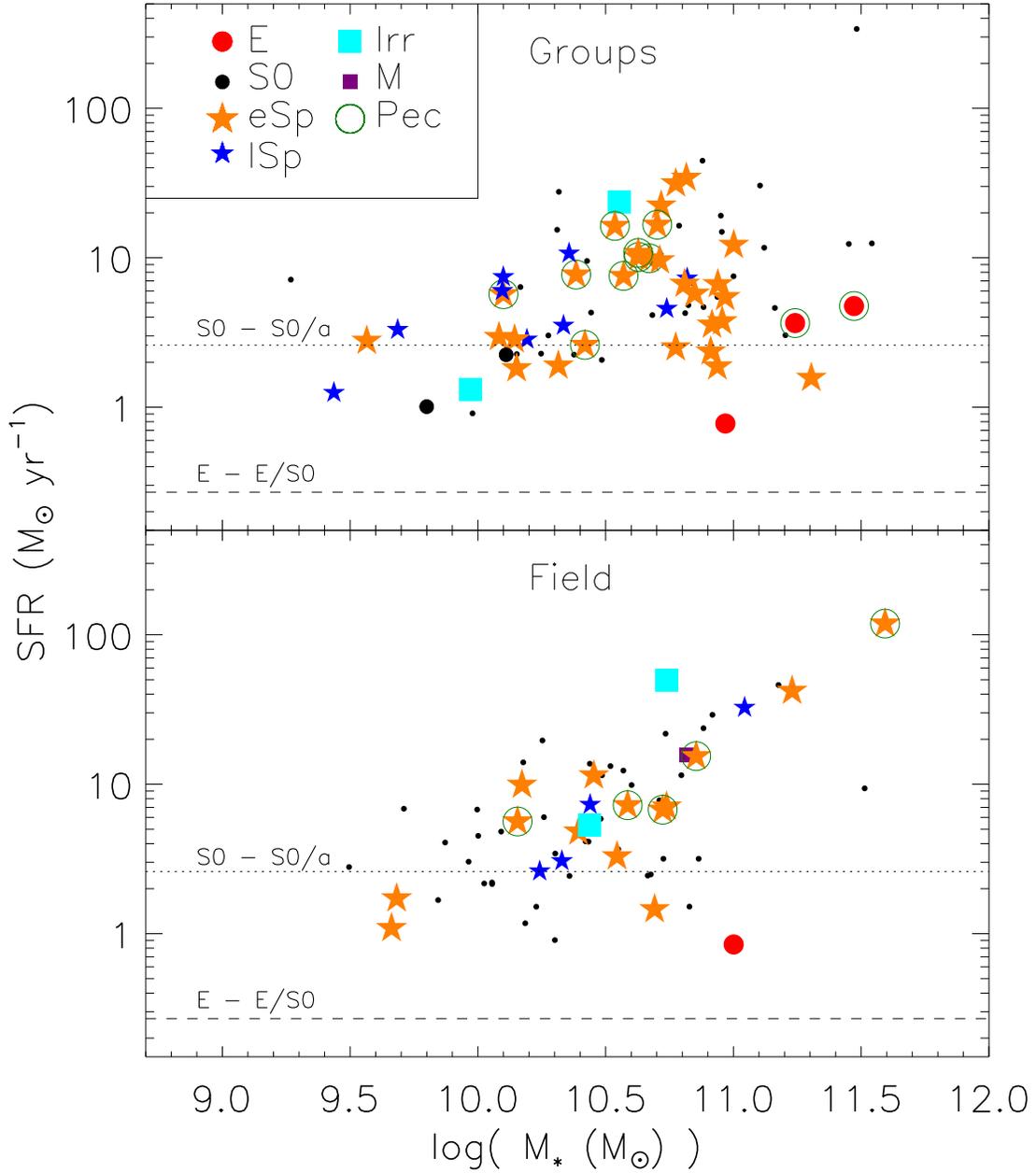}
\caption[]{SFR versus stellar mass for all galaxies detected at 24\micron~
(small black dots), separated by environment.  
Visually-classified morphologies, where available, are identified 
by different point styles and colors.
Ellipticals, S0s, and early-type
spirals tend to have higher masses than the other galaxy types, as expected.
The 99th percentile of 60\micron~luminosity
for E--E/S0 and S0--S0/a galaxies in the Nearby Galaxies Catalog \citep{tully89} 
correspond to 0.27 M$_{\odot}$ yr$^{-1}$ (dashed line) and 2.6 M$_{\odot}$ yr$^{-1}$
(dotted line), respectively.  All of the ellipticals detected at 24\micron~fall far 
above the SFR seen in local ellipticals, while the two S0s fall close
to the 99th percentile for local S0s.}
\label{fig:morph_env}
\end{center}
\end{figure*}

Thus, even by these stringent tests, 
we have detected four elliptical and two S0 galaxies at 
24\micron, none of which have an obvious
AGN contribution.  Five of these galaxies (three ellipticals and two S0s) are in 
groups.  We cannot tell whether the groups and field are different in this regard at a
statistically significant level.  However, among the IR-detected group members,
$\ga$ 6\% are early types.  That is, some early-type galaxies in groups were forming stars
at significant levels around z $\sim$ 0.5.  We can compare this behavior with that
of local early-type galaxies using the study of \citet{dh97}.  They extracted IRAS 
60\micron~detections of galaxies in the Nearby Galaxies Catalog (NBGC; Tully 1989),
which contains 2367 galaxies within 40 Mpc (H$_{0}$ $=$ 75 km s$^{-1}$ Mpc$^{-1}$).
After eliminating ones not covered by IRAS or compromised for other reasons,
2094 remained in this study, of which 1215 were detected by IRAS (including
22\% of the 151 ellipticals).  We determined the 99th percentile of 60\micron~luminosity
for the E--E/S0 and S0--S0/a categories and converted it to L$_{TIR}$ of 2.4 $\times$
10$^{9}$ L$_{\odot}$ and 2.3 $\times$ 10$^{10}$ L$_{\odot}$, respectively.
The corresponding SFRs are 0.27 and 2.6 M$_{\odot}$ yr$^{-1}$
\citep{rieke09} and are indicated in Figure \ref{fig:morph_env} by dashed and
dotted lines, respectively.  
Figure \ref{fig:morph_env} also shows stellar mass versus SFR
for all galaxies in our sample detected at 24\micron~(small black dots), separated
by environment, with visual morphologies indicated by a variety of symbols.
The SFRs for the two group S0 galaxies are at or just below the 99th percentile 
for local S0--S0/a galaxies, while the SFRs of all four E galaxies exceed the 99th 
percentile for local E--E/S0 galaxies by factors of 3--20, an 
unexpected result given that there are less than 100 early-type group members
in our sample.  

On a more general note, these measurements imply that not all early-type galaxies 
are dead at z $\sim$ 0.5; many are still in the process of forming stars, indicating
a larger amount of evolution with redshift than previously thought (e.g., Larson 1974;
Chiosi \& Carraro 2002).  While it does
appear that there are still some early-type galaxies forming stars in the local
universe \citep{temi09}, the ubiquity of IR-active early-types in group and field
environments remains largely unknown.
We do know, however, that star formation in E/S0 galaxies has been decreasing since
z $\sim$ 1.  \citet{kaviraj08} find that star formation in early-type galaxies
has decreased by $\sim$50\% since z $\sim$ 0.7, and they show that local
early-type galaxies, while largely quiescent now, have had spurts of star formation
in their recent past.  Thus, the enhancement of star-forming activity in early-type
galaxies at z $\sim$ 0.5 compared with the present epoch may apply generally for
field and group members.

We now discuss the fraction of 24\micron-detected
galaxies with respect to the full range of morphologies, separated by groups and
field.  As just discussed, for E/S0 galaxies, 5 out of 75 in groups were detected at
24\micron~(7\%) and 1 out of 22 in the field (5\%).  eSps behave similarly
in the two environments:  29 out of 47 group galaxies were
detected (62\%) compared with 14 out of 23 (61\%) in the field.  For lSps, 
9 out of 19 (47\%) are detected in groups and 4 out of 20 (20\%) 
in the field.  Although there is a hint of a larger incidence of late-type galaxies
with high SFRs detected in our group sample, there are no differences at a significant level.

\section{DISCUSSION}


There is strong agreement that the dense environments in the cores of clusters do 
have a large effect on their member galaxies in terms of star formation.
\citet{marcillac07}, \citet{patel09}, and \citet{koyama10} show that the densest regions 
of the clusters RXJ 1716.4+6708 and RXJ 0152.7-1357 strongly suppress star formation.
Using the CNOC1 cluster galaxy sample (0.2 $\la$ z $\la$ 0.55), \citet{balogh00} find
that the mean galaxy SFR decreases with decreasing radius from the center of the
cluster, and \citet{ellingson01} shows a decrease in the fraction of blue and
emission-line galaxies and an increase in the fraction of ellipticals as one 
approaches the cluster core.  Similarly, for the more distant cluster MS 1054-03
(z $\sim$ 0.8), \citet{bai07} observed that star formation in member galaxies near 
the core was substantially suppressed; \citet{vulcani10} find the same suppression
in cluster galaxies from 0.4 $\la$ z $\la$ 0.8 as compared with the field, though
the reduction in SFR was more modest in those clusters.
Locally (z $\la$ 0.3), we continue to see star-formation quenched or 
suppressed near cluster cores:  \citet{haines09} find suppression for 
star-forming cluster galaxies (defined by L$_{IR}$ $>$ 10$^{10}$ L$_{\odot}$), as
do Bai et al. (2006, 2009), who report that star-forming galaxies
(defined by SFR $\ga$ 0.2 M$_{\odot}$ yr$^{-1}$) are much less likely to be found
in the cores of local clusters Coma and A3266.  \citet{bk04} use kinematics of
different cluster galaxy morphologies to show that lSps are
more likely to be found at higher distances from the core.

Although the SFR suppression in cluster cores is clearly established, 
work to-date has not reached a firm conclusion
about the comparison of group galaxies with those in the field and in clusters
with respect to the IR regime.  
\citet{wilman08} find, from 8\micron~measurements of a subset of the GEEC
sample, that the SFR
is significantly suppressed in groups as compared with the field at z $\sim$ 0.4.
\cite{bai09b} use 24\micron~observations to show that local groups have somewhat
suppressed SFRs compared with the field, and somewhat elevated ones compared with
clusters (though they have a lower SFR limit than our study). 
At higher redshifts, \citet{marcillac08} use a large sample of galaxies
in the Extended Groth Strip to find that LIRGs and ULIRGs measured at
24\micron~do not preferentially exist in higher-density environments (including groups)
at z $\sim$ 0.9.  \citet{tran09} use 24\micron~data to 
find a substantially (four times) higher incidence
of active star-forming galaxies in groups compared with clusters at z $\sim$ 0.37.
They also find that the groups and field are similar in this regard for their
luminosity-limited sample.  \citet{patel09}
use 24\micron~measurements to find a progressive suppression of the SFR by an order 
of magnitude from the field to cluster core densities.

Other recent studies in optical bands show suppression of star formation in groups
due to different mechanisms.  
\citet{peng10} and \citet{kovac10} use Sloan Digital Sky Survey and zCOSMOS galaxies
at a variety of redshifts (up to z $\sim$ 1) 
and environments to show that the effects of environment
on individual galaxies are separate from the evolution and quenching
with increasing galaxy mass.  This ``mass quenching'' is the dominant effect at high
galaxy masses (M$_{*}$ $\ga$ 10$^{10.2}$ M$_{\odot}$), 
while other effects, like environmental quenching, dominate for low-mass
and satellite galaxies (M$_{*}$ $\la$ 10$^{10}$ M$_{\odot}$; Peng et al. 2010).
As mentioned previously, \citet{tran09} find comparable fractions of star-forming
galaxies in their supergroup and the field for a luminosity-limited sample, but
their mass-selected sample shows that the supergroup has about half the fraction
of star-forming galaxies as the field.

In light of the uncertainties in the effects of groups on the SFRs as reflected in the
{\it Spitzer} 8 and 24\micron~data, we have carried out a thorough study of the CNOC2 groups at
0.3 $<$ z $<$ 0.55, which are exceptionally well-characterized and have substantial 
amounts of ancillary data. The large sizes of this group sample and of the accompanying field 
sample also allow reasonably good statistics for our conclusions. We find that
the incidence of 24\micron~emission is virtually the same in these groups as in the field.  
This result agrees well with that of \citet{tran09} for their luminosity-selected
supergroup at z = 0.37 but has 
higher statistical weight because our field sample is significantly larger. This agreement 
is interesting because \citet{tran09} studied super-groups with large velocity 
dispersions and significant X-ray luminosities, while our groups generally have lower
velocity dispersions and no X-ray emission.

This similarity of IR properties appears to hold in detail, both in the forms of 
the fLFs and as characterized by L$^*$. The overall fLFs for groups and 
field are consistent with being drawn from the same 
distribution. \citet{tran09} found that L$^*$ for their field sample was  
$\sim$ 0.4 dex lower than for their groups. 
To look for this effect, we used our fLF Schechter function fits to 
determine L$^*$ for our groups.  L$^*$ for our field is marginally larger (again 
$\sim$0.4 dex) than the groups, though given the error bars on the values of L$^*$,
the field and group L$^*$ are comparable. 
It is possible that the difference between the two studies arises from the 
different environments---one large group-like structure for \citet{tran09}
versus our individual, smaller groups. However, given the statistical significance of 
the two studies, it is plausible that there is no significant overall difference 
in L$^*$ between the two environments. This emphasizes the lack of any strong 
dependence of group member properties on the size or mass of the group 
(Wilman et al. 2005b; this work).

The IR properties of our group and field galaxies appear to be 
contrary to the shift in the distribution of galaxy morphologies
toward early-types in groups (McGee et al. 2008; Wilman et al. 2009), which we confirm with 
a form of CA analysis. In part, this apparent contradiction can be explained by the presence 
of a number of E/S0 galaxies that are detected at 24\micron; this IR activity appears to 
arise from elevated levels of star formation as compared with local E/S0 galaxies (as 
also found by \citet{tran09}). However, this behavior has also been seen at similar 
redshifts in field early-type galaxies (e.g., van der Wel et al. 2007; 
P\'{e}rez-Gonz\'{a}lez et al. 2008). Along with a higher population of E/S0 galaxies, 
the groups also have larger overall numbers of massive, 
eSps. These galaxies contribute significantly to the numbers of IR-detected 
group members and to the group LF. \citet{tran09} find a similar result 
for their super-group: that an excess population (compared with the field) of spirals fills 
in the dearth of star formation that would otherwise exist because of the larger 
proportion of early-type galaxies.  A minor difference is that the \citet{tran09}
group sample is rich in relatively low-mass star-forming spirals compared with those
in our groups.

An interesting observation is that there are only two galaxies
total in our group and field sample that can be classified as ULIRGs 
(12 $\la$ log(L$_{TIR}$ (L$_\odot$)) $\la$ 13).  \citet{lefloch05}
showed that ULIRGs make up only $\sim$10\% of the IR population at z $\sim$ 0.7,
which roughly means, at the redshifts we are studying here, we should expect to see
4--12 ULIRGs.  While the errors on this estimate are almost certainly large
enough to encompass our two ULIRGs, it is still odd that our sample lies at the
low end of the expected range.  If galaxies in groups have higher propensities for
mergers or low-velocity encounters, we might expect to see more of these
high-L$_{TIR}$ galaxies in the groups, if not overall.  \citet{geach09}
suggest that such encounters could be fueling the growth of galactic bulges, which
explains the larger fraction of Sa through E galaxies in the groups.  If 
we do not see massive amounts of star formation in the groups---or, in this case,
neither the groups nor the field---then either a process common to both group
and field galaxies is responsible for the bulge formation or 
star formation from mergers and low-velocity encounters is short enough to not
show significant IR emission in individual galaxies.

Because mass has been shown to have significant association with suppressing star formation,
we also made a quantitative test of the overall similarity of group and field members 
of similar mass. The ratio of low-activity star forming galaxies (i.e., those with 
SFR $<$ 2.7 M$_\odot$  yr$^{-1}$) to the total number of galaxies with 
M$_{*}$ $>$ 10$^{10}$ M$_\odot$ is 0.69.  We synthesized this result from field 
galaxies in a Monte Carlo calculation, drawing randomly from a field sample matched in mass. 
The synthesized distribution is Gaussian and has its maximum probability at a fraction of 
0.66, with a range at $\pm$1$\sigma$ from 0.62 to 0.70. That is, there is only
marginal evidence (1$\sigma$) for a change in incidence of high levels of star formation 
in galaxies of the same mass in groups versus the field, and any such change is limited 
to be no more than a 10\% effect (at 1$\sigma$).

All of these results are consistent with the hypothesis that the difference between 
the group and field populations is confined largely to their differing mass functions. 
For a given mass and morphological type, there is no statistically significant suppression 
or enhancement of star formation in individual group galaxies in the intermediate stellar 
mass range of 10$^{10}$ to 2 $\times$ 10$^{11}$ M$_{\odot}$. However, groups have begun to 
build massive galaxies with lower sSFRs as typical for their 
relatively-early types. The shifts toward lower sSFRs and toward 
higher masses tend to cancel each other, leading to similar infrared fLFs.

We must remind ourselves, however, that we are only probing the brightest star-forming
galaxies at these redshifts; galaxies with SFRs below our detection limit
could be more affected by the group environment, resulting in a lower
fraction of star-forming galaxies in the groups \citep{peng10}.
Locally, the fraction of IR-active galaxies in the two
environments differs by $\sim$30\% (more star-forming galaxies in the field) for
galaxies with SFRs $\ga$ 0.1 M$_{\odot}$ yr$^{-1}$, a much lower detection limit
than our sample \citep{bai09b}. Other group
studies have found environmental dependence of the fractions of star-forming
group and field galaxies at fixed luminosity or stellar mass using
different indicators (Wilman et al. 2005a, 2008; Balogh et al. 2007, 2009).


These differences suggest that groups are indeed an intermediate stage between the 
field and clusters. Our groups contain
fractions of E and S0 galaxies at levels comparable to clusters (Wilman et al. 2009), and
the mass distribution of group galaxies tends to extend higher than that of field 
galaxies, as confirmed with a more detailed inspection of the group and field galaxy 
masses. Despite these differences, the overall IR activity in 
groups seems to indicate a lack of suppression or enhancement of star formation as 
compared with the field: the fractions of star-forming galaxies 
(SFR $>$ 2.7 M$_\odot$ yr$^{-1}$) are comparable in the groups and field,
and the fLFs are nearly identical. Individual galaxies of 
similar mass and morphology appear to have virtually identical infrared properties in 
the two environments. Thus, the group environment affects
the masses and morphologies of galaxies, and their star forming properties change 
consistent with these effects. However, any additional changes in star forming properties 
are, at best, subtle, at least for
SFR $>$ 2.7 M$_{\odot}$ yr$^{-1}$, indicating that the level of star formation is driven 
primarily by galaxy mass, itself a function of environment. In other words, star-forming 
activity in individual galaxies is only indirectly related to the group versus field 
environment but is linked more strongly to the overall change in galaxy masses and 
morphologies. The higher L$^*$ for the supergroup of \citet{tran09} may indicate 
an enhancement of star formation that is environmentally-dependent, but they find that
mass more strongly affects star formation.  Apparently, the assumed
increased rate of galaxy-galaxy interactions in groups either does not affect the star 
formation significantly, or strong interaction-driven star formation occurs in environments
other than the groups we studied here.  

The outskirts of clusters are a probable alternative
location for galaxy processing, as shown by the higher fractions of IR-bright galaxies
in group and field galaxies than the outskirts of the Coma and A3266 clusters \citep{bai09b}.
Additional studies of groups and clusters at a variety of redshifts, preferably
to lower SFR limits, are needed to further disentangle the effects of these moderately-dense 
environments on the star formation, mass, and morphology of their member galaxies.

\section{CONCLUSIONS}

We have observed 26 galaxy groups and accompanying field galaxies with deep
MIPS photometry and used 24\micron~flux densities to estimate the total IR luminosities
and SFRs of galaxies in both environments.  We find that on an individual
basis, group and field galaxies of similar mass and morphology 
do not differ significantly in terms of their
SFRs, and the amount of star formation does not depend on the richness of the
groups.  However, the groups have systematically lower sSFRs and  
a higher incidence of massive
early-type galaxies, more reminiscent of clusters than the field.  We discovered
that some of these E and S0 galaxies, as well as a large contingent of massive early spirals,
are still forming stars at significant levels.  These galaxies may explain why
the fLFs of the groups and field are nearly identical
despite the overall decrease in star-forming activity in the groups.  
The group environment affects galaxy SFRs primarily through the shift toward higher masses,
with an accompanying trend toward earlier types and reduced sSFRs.
These high-mass, early-type galaxies, along with IR luminosities comparable to the field, 
put groups in between the field and clusters in terms of overall galaxy properties.



\acknowledgements

We thank Michael Balogh and Ann Zabludoff for helpful input, the referee
for his/her comments and suggestions, and 
the CNOC2 team for access to their unpublished data.
LCP acknowledges financial support from an NSERC Discovery grant.
We thank NASA for financial support through contract 1255094 from Caltech/JPL
to the University of Arizona.
This work is based on observations made with the {\it Spitzer Space Telescope},
which is operated by the Jet Propulsion Laboratory, the California Institute of Technology,
and NASA.   

\newpage

\clearpage
\section*{Appendix}

\begin{figure*}
\begin{center}
\centering
\includegraphics[width=15cm]{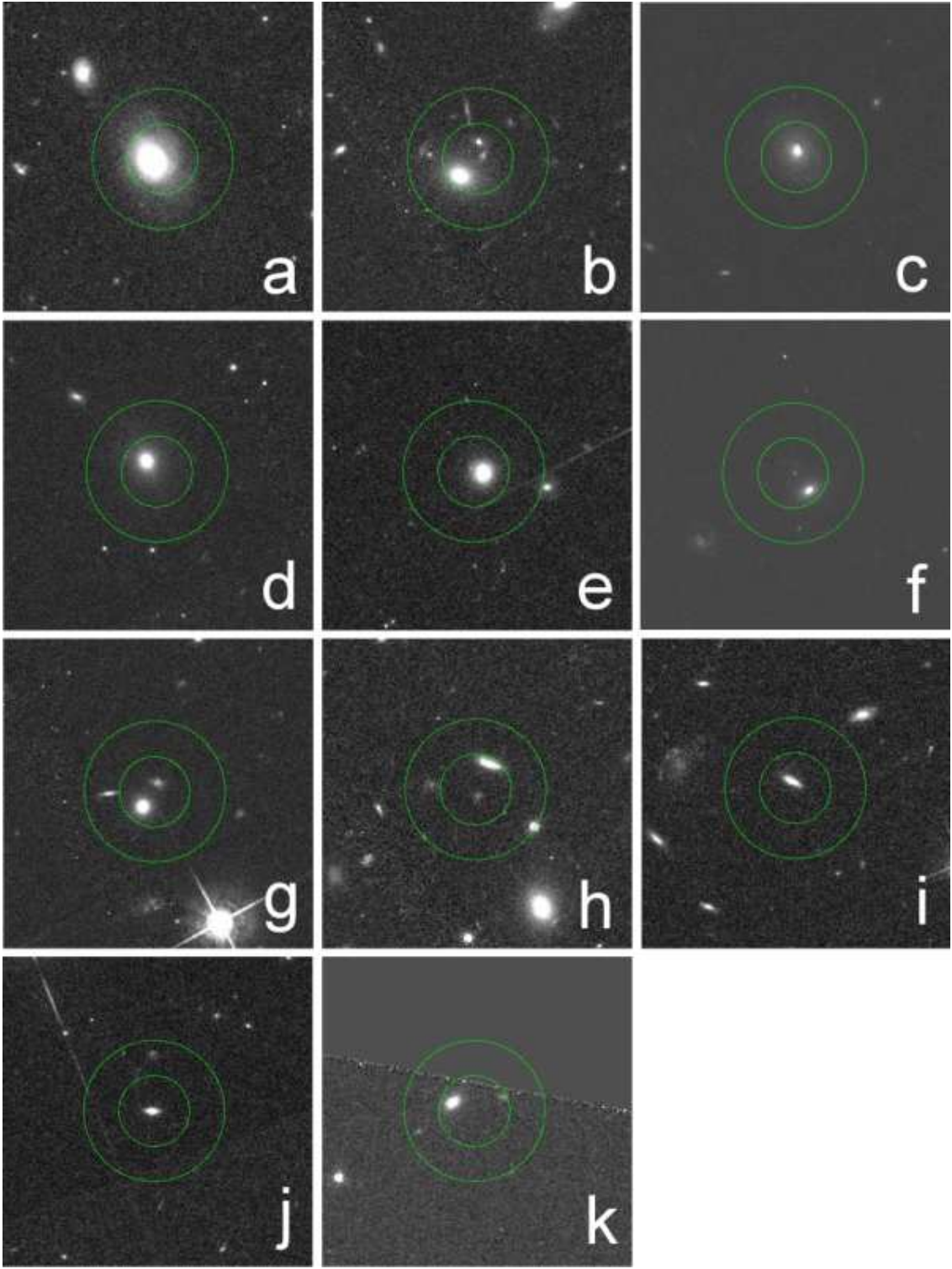}
\caption[]{HST ACS images of the 11 E/S0 galaxies matched with 24\micron~sources.
The inner annulus has a radius of 3 arcsec, which is the length we used for matching
the optical coordinates with 24\micron~sources.  The outer annulus shows the FWHM
of the MIPS PSF.  The ID and patch number of the matched galaxies are as follows:  (a) 
1447 040969, (b) 1447 020364, (c) 1447 041307, (d) 1447 111706, (e) 1447 122388, (f)
1447 150408, (g) 1447 111547, (h) 1447 120982, (i) 1447 091003, (j) 1447 091304, and (k)
2148 141211.}
\label{fig:hst_acs}
\end{center}
\end{figure*}

Our initial coordinate matching of MIPS 24\micron~sources to their optical counterparts
resulted in 11 early-type (E or S0) galaxies with possible IR emission:  seven group
and four field galaxies (Table 1). Given the density of galaxies in our fields---and how
surprising it was to find E/S0 galaxies with possible star formation---we checked
each match individually to confirm whether or not the 24\micron~emission was 
coming from the early-type galaxy or if there were a possibility that a close 
companion was the actual source of the emission.  Here, we provide a short description 
of both the optical and 24\micron~sources, as well as HST ACS postage stamps of each
galaxy (Figure \ref{fig:hst_acs}) with error circles matched to the 
24\micron~emission at 3 arcsec 
(matching radius) and 6 arcsec (radius of MIPS PSF FWHM).

PATCH 1447 ID 40969 (Figure \ref{fig:hst_acs}a):  While classified as an ``E pec'' galaxy (elliptical
with a possible interaction from a companion), there are no sources within the 3 arcsec
matching radius, and the 24\micron~emission is clearly right on top of the elliptical
(the emission peaks less than 1 arcsec from the center of the elliptical).
The 24\micron~emission is definitely coming from the early-type galaxy with no other
obvious source.  

PATCH 1447 ID 020364 (Figure \ref{fig:hst_acs}b):  This elliptical galaxy has three small neighbors,
though only two are within the 3 arcsec matching radius (the others are slightly farther
out).  The 24\micron~source is closer to the nearest neighbor and close to the edge 
of the 3 arcsec radius.  It is possible that the IR emission is coming from the companion
galaxy and not the early-type, and so it is not included in our list of IR-active E/S0
galaxies.

PATCH 1447 ID 041307 (Figure \ref{fig:hst_acs}c):  This galaxy is another peculiar elliptical but with no
galaxies within 3 arcsec.  The 24\micron~source, while faint, is almost directly
on top of the elliptical ($\sim$1 arcsec away).  
It is highly unlikely that the IR emission is coming from another
source, so we classify this galaxy as an IR-active early-type.

PATCH 1447 ID 111706 (Figure \ref{fig:hst_acs}d):  Classified as a normal elliptical galaxy, this source
has faint 24\micron~emission that lies almost on top of the galaxy ($\sim$1 arcsec) with no other
obvious galaxies within the match radius.  This object has deep X-ray coverage as well,
with no detected X-ray source to account for the IR emission.  
We include this elliptical in our list of star-forming early-type galaxies.

PATCH 1447 ID 122388 (Figure \ref{fig:hst_acs}e):  Other than being classified as an ``E/S0,''
(somewhere between an elliptical and an S0 galaxy, but more closely resembling an
elliptical) this galaxy's situation is almost identical to the previous one.

PATCH 1447 ID 150408 (Figure \ref{fig:hst_acs}f):  Here is another regular elliptical galaxy with only one
point-source-like neighbor within 3 arcsec.  The 24\micron~source is well within
the match radius, but it is between the nearby object and the elliptical ($\sim$2 arcsec
from the elliptical).  It is
uncertain as to which source the 24\micron~emission is coming from, so we removed
this galaxy from being a star-forming early-type.

PATCH 1447 ID 111547 (Figure \ref{fig:hst_acs}g):  This is a normal elliptical galaxy with two close
companions, one of which appears to be a faint irregular fully within the matching
radius.  The MIPS source appears extended and spans the distance between this
closer neighbor and the elliptical.  As with the previous galaxy, the uncertainty
in the source of the 24\micron~emission forced us to not include this galaxy as
an IR-active source.

PATCH 1447 ID 120982 (Figure \ref{fig:hst_acs}h):  A very faint irregular galaxy barely lies within
3 arcsec of the early-type galaxy, listed as a peculiar S0.  The IR source is again
between the two galaxies, and though it is slightly closer to the faint companion,
the 24\micron~emission is well within the match radius boundary ($\sim$2 arcsec from the
S0).  We conservatively do not include this early-type as IR-active.

PATCH 1447 ID 091003 (Figure \ref{fig:hst_acs}i):  This galaxy is an S0/a, meaning it closely resembles
an S0 galaxy but with some Sa qualities.  There are no other galaxies within
3 arcsec, and the IR source peaks $\sim$1 arcsec from the S0/a.  We include
this object as a star-forming early-type galaxy.

PATCH 1447 ID 091304 (Figure \ref{fig:hst_acs}j):  This normal S0 is almost identical to the previous
galaxy.

PATCH 1447 ID 141211 (Figure \ref{fig:hst_acs}k):  As with the previous two galaxies, this S0/E (S0
galaxy somewhat similar to an elliptical) has no other companions within 3 arcsec (though
there are two faint galaxies between 3.5 and 5 arcsec away).  The
24\micron~emission is slightly offset from the galaxy ($\sim$1 arcsec) but within the match 
radius.  A very faint galaxy lies $\sim$1.5 arcsec from the 24\micron~position.
While it seems likely that the IR emission is coming from the S0 galaxy,  
we took a conservative stance and do not classify this galaxy as IR-active.

\begin{figure*}
\begin{center}
\centering
\includegraphics[width=15cm]{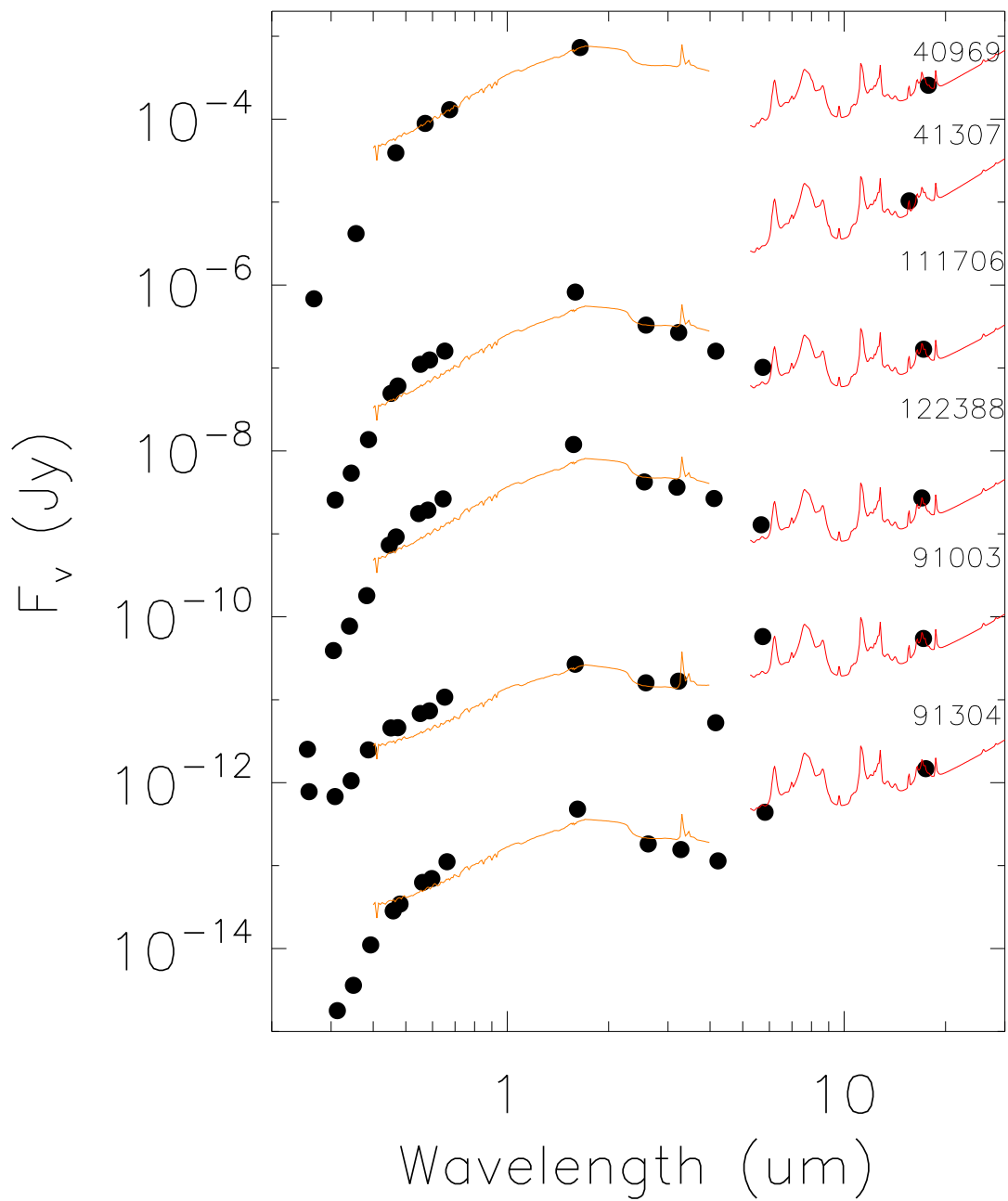}
\caption[]{Optical and IR SEDs (rest-frame) for all six star-forming E/S0 galaxies, identified
by their GEEC ID and fit by different segments of the \citet{rieke09} average
star-forming galaxy templates.  Each galaxy was matched to the average
template with the closest L$_{TIR}$ and fit to the 24\micron~data point (red 
spectra).  The orange spectra are best-fit ($\chi^{2}$) average star-forming templates 
for the optical/near-IR photometry; while these fits are not very accurate, they
are sufficient for our purposes.  Except for 40969 and 41307, all galaxies
have enough photometric points to indicate the presence of a stellar bump and
mid-IR dust emission similar to normal star-forming galaxies.}
\label{fig:seds}
\end{center}
\end{figure*}

In Figure \ref{fig:seds}, we plot rest-frame optical and IR SEDS for the 
six E/S0 galaxies that have 
confirmed IR emission.  The stellar outputs of these galaxies may be dominated by
a relatively old population.  Therefore, we fit \citet{rieke09} average 
star-forming galaxy templates
to the SEDs in two distinct segments.  First, we found the average template with 
the closest L$_{TIR}$ to each galaxy and fit the 24\micron~data point to the
template (red spectrum).  For the normal stellar emission, we used optical and near-IR
photometry to find the best-fit ($\chi^{2}$) average star-forming template (orange
spectrum).  Except for the first two galaxies, we see a clear stellar
bump from normal stars and then increasing luminosity of the mid-IR dust emission
beyond 7\micron, indicative of a normal star-forming galaxy and not an AGN.  
In all cases, the SED is consistent with the expectation for star formation.  There 
are two cases where 1) there is IRAC data; and 2) a detection of the aromatic emission
at 8\micron~is predicted from the combination of IRAC bands 1--3 and the MIPS 
24\micron~flux density.  The expected 8\micron~excess is seen for both galaxies.

\begin{deluxetable}{lccccccccc}
\tablecolumns{10}
\tablecaption{Early-Type Galaxies Matched with IR Sources}
\rotate
\label{table:earlytypes}
\tablehead{
\colhead{Patch\tablenotemark{a}} &
\colhead{ID} &
\colhead{R.A.} &
\colhead{Decl.} &
\colhead{z} &
\colhead{L$_{TIR}$} &
\colhead{SFR} &
\colhead{Group\tablenotemark{b}} &
\colhead{Galaxy} &
\colhead{X-ray} \\
\colhead{} &
\colhead{} &
\colhead{} &
\colhead{} &
\colhead{} &
\colhead{(10$^{10}$ L$_{\odot}$)} &
\colhead{(M$_{\odot}$ yr$^{-1}$)} &
\colhead{} &
\colhead{Type} &
\colhead{Coverage?}
}
\startdata
1447 & 040969 & 222.377813 & 9.511117 & 0.35 & 3.8  & 3.7  & 23 & E pec  & N \\
1447 & 020364 & 222.431575 & 9.232753 & 0.36 & 11   & 12   & 25 & E      & N \\
1447 & 041307 & 222.427158 & 9.536456 & 0.54 & 4.9  & 4.8  & 39 & E pec  & N \\
1447 & 111706 & 222.483754 & 8.940261 & 0.39 & 0.89 & 0.78 & 32 & E      & Y \\
1447 & 122388 & 222.355163 & 9.074511 & 0.41 & 0.96 & 0.85 &  0 & E/S0   & Y \\
1447 & 150408 & 222.184342 & 8.842967 & 0.32 & 1.7  & 1.5  &  0 & E      & N \\
1447 & 111547 & 222.470204 & 8.926272 & 0.30 & 1.0  & 0.91 &  0 & E      & Y \\
1447 & 120982 & 222.349458 & 8.990908 & 0.37 & 3.0  & 2.8  &  0 & S0 pec & Y \\
1447 & 091003 & 222.578538 & 9.010039 & 0.40 & 1.1  & 1.0  & 32 & S0/a   & N \\
1447 & 091304 & 222.578025 & 9.030156 & 0.37 & 2.4  & 2.2  & 28 & S0     & N \\
2148 & 141211 & 327.600550 & -5.681283 & 0.44 & 4.5 & 4.3  & 138 & S0/E  & Y \\
\enddata
\tablenotetext{a}{Original CNOC2 patch number \citep{yee00}.}
\tablenotetext{b}{Galaxy group to which the object belongs; a 0 value indicates a field galaxy.}
\end{deluxetable}

\end{document}